\documentclass[12pt]{article}
\usepackage[margin=25mm]{geometry}
\usepackage{amsmath}
\usepackage{graphicx,psfrag,epsf}
\usepackage{enumerate}
\usepackage{natbib}
\usepackage{amssymb}
\usepackage{epstopdf}
\usepackage{enumerate}
\usepackage{placeins}
\usepackage{setspace}
\usepackage{float}
\usepackage{soul,color}
\usepackage{placeins}
\usepackage{upgreek, mathtools,bm}
\usepackage{algorithm,algpseudocode,tabularx}
\usepackage{caption}
\usepackage{subcaption}
\usepackage{enumitem}
\usepackage{multirow}
\usepackage{booktabs}
\usepackage{amsfonts,amssymb,graphics,amsthm, comment}
\usepackage{url}
\RequirePackage[colorlinks,breaklinks,linkcolor=blue,citecolor=blue,urlcolor=blue]{hyperref}
\usepackage{multicol}
\usepackage{listings} 
\usepackage{xcolor} 
\immediate\write18{texcount -tex -sum  \jobname.tex > \jobname.wordcount.tex}

\graphicspath{{./plots/}}
\newtheorem{theorem}{Theorem}[section]

\newtheorem{corollary}[theorem]{Corollary}
\newtheorem{proposition}[theorem]{Proposition}

\theoremstyle{remark}

\theoremstyle{plain}
\theoremstyle{definition}
\newtheorem{remark}{{\bf Remark}}
\newtheorem{assumption}{{\bf Assumption}}
\newcounter{scenario}

\providecommand{\keywords}[1]
{
  \small	
  \textbf{\textit{Keywords---}} #1
}

\title{Estimation and Inference for One-pass Streaming Function-on-Function Regression}
\author{Shuyang Jiang$^{1}$, Xin Liu$^{1}$, and Peijun Sang$^{2}$ \\
        \small $^{1}$Department of Statistics and Data Science, Shanghai University of Finance and Economics \\
        \small $^{2}$Department of Statistics and Actuarial Science, University of Waterloo \\
}
\date{} 

\begin{document}
\maketitle

\begin{abstract}
In many applications, functional pairs are collected sequentially from a data stream. Existing function-on-function regression methods are primarily designed for batch data, and bootstrap inference has become one of the most popular approaches. However, this procedure may be infeasible when observations cannot be stored. In this paper, we propose a one-pass functional SGD estimator that updates upon each observation's arrival without data storage, and a weighted rolling validation rule is provided for step size selection for implementation. Further, an online multiplier bootstrap is developed for inference that updates perturbed stochastic gradient paths in parallel, bypassing estimating an infinite-dimensional covariance operator.
Asymptotic behaviors are established for the proposed method, and conditional bootstrap validity is justified for the bias-centered pointwise estimation error.
Simulation studies and two real data applications confirm the empirical effectiveness of the proposed procedures. To the best of our knowledge, this is the first theoretically justified one-pass framework for bootstrap inference on a bivariate coefficient surface with streaming functional data.
\end{abstract} \hspace{10pt}


\section{Introduction}\label{sec:intro}
Modern scientific studies increasingly collect observations over continuous domains, giving rise to different forms of functional data, such as curves or trajectories, and functional data analysis provides a systematic framework for studying such observations. One of the most powerful methods is the functional regression that characterizes statistical relationships among functionals; see, for example, the monographs \citep{Ramsay1997,hsing2015theoretical,kokoszka2017introduction} and the survey paper \citep{wang2016functional}.

A substantial literature in functional regression has concentrated on the scalar-on-function model, in which a scalar response is associated with one or several functional predictors. To regularize the slope function, \citet{CardotFerratySarda2003} proposed a penalized B-spline estimator. Framing slope estimation as an ill-posed inverse problem, \citet{CaiHall2006} clarified the fundamental distinction between prediction and slope recovery, and optimal minimax rates under this perspective were established for functional principal component (FPC) and quadratic regularization methods \citep{HallHorowitz2007}. 
Furthermore, reproducing kernel Hilbert space (RKHS) methods were proposed for estimation and prediction, and their optimality properties were established \citep{yuan2010reproducing,Cai01092012}.

In parallel with these developments, 
function-on-function models were introduced to relate functional predictors directly to functional responses. Specifically,  \citet{RamsayDalzell1991} formulated a general functional linear model.
This function-on-function setting extends beyond mere notation, requiring estimation to regularize a bivariate slope surface \(\beta_0(t,s)\), accommodate functional errors, and distinguish response prediction from surface recovery.
Existing methods addressing these challenges have been developed predominantly for batch data, following two main paradigms. The first relies on dimension reduction, including FPC expansions, which have been used to study estimation and prediction for this model with sparsely and irregularly observed trajectories contaminated with measurement error \citep{yao2005functional}. Subsequent work further established optimal prediction rates for fully observed trajectories \citep{CrambesMas2013} and minimax rates for surface recovery \citep{imaizumi2018pca}.
However, since FPCs capture only the marginal variation of the predictor and response, they may not efficiently represent the bivariate regression surface \citep{sun2018optimal}.
An alternative is to regularize the coefficient surface or regression operator directly. Prominent examples include the RKHS framework \citep{LIAN2015395} and the penalized signal compression approach \citep{LuoQi2017}.
In particular, a representer theorem and minimax optimality for mean prediction were established under a tensor-product RKHS formulation \citep{sun2018optimal}, while this minimax optimality was further extended to arbitrary sampling schemes ranging from sparse to dense designs \citep{yan2026functional}.
On the inference front, a batch inference procedure was developed to yield simultaneous confidence regions for \(\beta_0(t, s)\) and prediction bands for the functional response \citep{dette2024statistical}.
Despite theoretical guarantees, these methods all require full-sample access and thus may not be appropriate for the one-pass streaming scenario, when functional observations are collected sequentially.

Indeed, in the context of streaming functional data, recomputing empirical covariance or Gram operators to solve global regularized problems upon each new arrival is computationally prohibitive, while storing an ever-growing collection of trajectories is often infeasible. 
Consequently, stochastic approximation \citep{robbins1951stochastic}, coupled with Polyak--Ruppert iterate averaging \citep{ruppert1988efficient,polyak1992acceleration}, provides a natural paradigm for processing functional streams sequentially.
In nonparametric least-squares problems, averaged stochastic gradient descent (SGD) also serves as an implicit regularization mechanism, enjoying sharp convergence guarantees in general Hilbert and RKHS settings \citep{bach2013non, dieuleveut2016nonparametric, mucke2019beating}. For scalar-on-function linear regression, \citet{ChenTangFanGuo2022} developed RKHS-based online gradient descent algorithms and established prediction rates under online and finite-horizon step size schedules. These results confirm the feasibility of learning from streaming functional predictors but do not address functional responses, bivariate coefficient surfaces, or online statistical inference. 

Beyond point estimation, online uncertainty quantification poses a distinct challenge to online learning. Conventional bootstrap procedures require repeatedly resampling the stored dataset and thus fail when historical observations are discarded, whereas plug-in asymptotic inference is hindered by the need for estimating an infinite-dimensional covariance structure \cite[Theorem 3.2]{dette2024statistical}.
For parametric stochastic approximation, \citet{fang2018online} introduced a perturbation-based online bootstrap that updates randomly weighted gradient trajectories alongside the original estimator. 
This paradigm was subsequently extended to robust online estimation and bootstrap inference in function-on-scalar regression \citep{cheng2025online} and to one-pass estimation and online bootstrap inference for a univariate slope function in scalar-on-function regression \citep{xie2026scalable}.
Although these works establish the viability of online inference for either the functional predictors or the functional response, neither encompasses the function-on-function regime, where the target is a bivariate surface coupled with a functional error process. Together with the batch-only nature of existing function-on-function procedures, this leaves a fundamental gap in one-pass estimation and inference for a bivariate coefficient surface. 

Against this background, we develop a one-pass framework for estimation and pointwise inference in streaming function-on-function linear regression. Our main contributions are threefold. 
First, the proposed RKHS-based functional SGD estimator can update the bivariate coefficient surface upon each incoming functional pair without storing historical observations, complemented by weighted rolling validation for step size selection.
Second, explicit convergence rates are established for both prediction error and estimation error of the coefficient surface, matching the  offline minimax rates for prediction \citep{sun2018optimal,dette2024statistical} and estimation \citep{imaizumi2018pca}, respectively.
Third, a scalable and easily implementable online multiplier bootstrap for pointwise inference is developed, whose bootstrap recursions can be updated in parallel, facilitating efficient computation while preserving the streaming paradigm. Alongside this practical utility, we establish its conditional consistency for the bias-centered pointwise error and the true coefficient surface subject to an additional bias-negligibility assumption.
To the best of our knowledge, this is the first statistically sound one-pass framework for performing inference for function-on-function linear regression with streaming functional data. 
Beyond theoretical guarantees, comprehensive simulation studies and real data applications to ERA5 wind--wave fields and Chapman--Shaoxing electrocardiogram recordings demonstrate the finite-sample performance of the proposed procedures.

The remainder of the paper is organized as follows. Section~\ref{sec:methodology} formulates the streaming function-on-function regression model and
develops the one-pass estimation and inference methodology, and their asymptotic convergence rates and pointwise bootstrap validity are established in Section~\ref{sec:theory}. Sections~\ref{sec:simulation} and~\ref{sec:realdata} report simulation studies and real data applications, respectively, and Section~\ref{sec:conclusion} concludes. Additional technical details, numerical studies, and theoretical proofs are deferred to the Supplementary Materials.

\section{Methodology}\label{sec:methodology}

We first introduce the function-on-function linear regression model for streaming data within a tensor-product RKHS framework in Section~\ref{subsec:model-setup}, and then propose a one-pass functional SGD estimator in Section~\ref{subsec:online_SGD}. An online multiplier bootstrap procedure is developed for pointwise inference in Section~\ref{subsec:online-bootstrap}. For practical implementation, a weighted rolling validation (wRV) criterion is presented in Section~\ref{subsec:adaptive-step} to select the SGD step size.

\subsection{Model and problem setup}\label{subsec:model-setup}

To start with, let \(\{D_i=(X_i,Y_i)\}_{i\geq1}\) denote sequentially arriving i.i.d. functional pairs, where \(X_i\in L^2(\mathcal{S})\) and \(Y_i\in L^2(\mathcal{T})\) denote the centered functional predictor and response on compact intervals \(\mathcal{S}\) and \(\mathcal{T}\subset \mathcal{R}\), respectively. Without loss of generality, we take \(\mathcal{T}=\mathcal{S}=[0,1]\). To capture their statistical association, a function-on-function linear regression model is considered  \citep{cuevas2002linear,yao2005functional}
\begin{equation}\label{eq:base_model}
    Y_i(t) = \int_{\mathcal{S}} X_i(s)\beta_0(t,s)\,ds + \varepsilon_i(t),
    \qquad t\in\mathcal{T},
\end{equation}
where \(\beta_0\) is the unknown bivariate slope surface, and \(\varepsilon_i\in L^2(\mathcal{T})\) is a centered random error process independent of \(X_i\), with its variance \(\sigma^2:=\mathbb{E}\|\varepsilon_i\|_{L^2(\mathcal{T})}^2<\infty\). Given the data stream \(\mathcal{D}_n=\{D_i:1\leq i\leq n\}\) observed up to the \(n\)-th step, our goal is to estimate \(\beta_0\) and conduct pointwise inference on \(\beta_0(t,s)\) in a strictly one-pass manner, i.e., updating the estimates upon the arrival of each \(D_i\) without retaining past data.

To regularize surface smoothness and support pointwise evaluation, we focus on the RKHS approach to functional linear regression \citep{ sun2018optimal}, and model \(\beta_0\) within a reproducing kernel framework. Recall that a separable Hilbert space \(\mathcal{H}_{\mathcal{X}}\) of real-valued functions on a compact domain \(\mathcal{X}\) is an RKHS if the point evaluation functional \(\delta_x: \mathcal{H}_{\mathcal{X}} \to \mathbb{R}\), \(\delta_x(f) = f(x)\), is continuous for every \(x \in \mathcal{X}\) \citep{cucker2002mathematical}. When \(\mathcal{X}\) is a compact interval, a canonical example is the Sobolev space of an integer order \(m \ge 1\)
\[
H^m(\mathcal{X}) = \left\{ f:\mathcal{X}\to\mathbb{R} \,\middle|\, f^{(j)} \text{ is absolutely continuous for } j=0,\ldots,m-1, f^{(m)}\in L^2(\mathcal{X}) \right\},
\]
and it is an RKHS when endowed with the squared norm
$$
\|f\|^2 = \sum_{j = 0}^{m - 1} \left\{\int_{\mathcal X} f^{(j)}(x) dx\right\}^2 + \int_{\mathcal X}  \{f^{(m)}(x)\}^2 dx;
$$
see Chapter 2.3 of \cite{gu2013}.
In the sequel, 
we take the marginal spaces \(\mathcal{H}_{\mathcal{T}}\) and \(\mathcal{H}_{\mathcal{S}}\) defined on \(\mathcal{T}\) and \(\mathcal{S}\) as the Sobolev RKHS \(H^m([0,1])\) of the $m$-th order, corresponding to the reproducing kernels \(K_{\mathcal{T}}\) and \(K_{\mathcal{S}}\), respectively. We assume that the slope surface \(\beta_0\) resides in the tensor-product RKHS \(\mathcal{H} := \mathcal{H}_{\mathcal{T}} \otimes \mathcal{H}_{\mathcal{S}}\) defined on \(\mathcal{T}\times\mathcal{S}\), whose reproducing kernel is \(K\bigl((t,s),(t',s')\bigr) = K_{\mathcal{T}}(t,t')K_{\mathcal{S}}(s,s')\) for  \((t,s)\) and \((t',s')\in\mathcal{T}\times\mathcal{S}\). The same setup was considered in \cite{sun2018optimal}. 
Let \(K_z=K(z,\cdot)\in\mathcal{H}\) for any $z = (t, s) \in \mathcal T \times \mathcal S$, and the reproducing property of $K$ immediately yields
 \(f(z)=\langle f,K_z\rangle_{\mathcal{H}}\) for all \(f\in\mathcal{H}\).

\subsection{An online functional stochastic gradient descent estimator}\label{subsec:online_SGD}

To fit the function-on-function regression model in \eqref{eq:base_model} given \(\mathcal{D}_n\), we consider minimizing the empirical integrated squared error criterion
\[
    \ell_n(\beta) = \frac{1}{2n} \sum_{i=1}^{n} \int_{\mathcal{T}} \left\{ Y_i(t) - \int_{\mathcal{S}}X_i(s)\beta(t,s)\,ds \right\}^{2}\,dt,
    \qquad
    \beta\in\mathcal{H},
\]
with respect to $\beta$. For each incoming predictor \(X_i\), define
\begin{equation}\label{eq:eta_Psi_definition}
    \eta_i = \int_{\mathcal{S}} X_i(s)K_{\mathcal{S}}(s,\cdot)\,ds \in \mathcal{H}_{\mathcal{S}} \quad \text{and} \quad \Psi_{i,t} = K_{\mathcal{T}}(t,\cdot)\eta_i \in \mathcal{H},
\end{equation}
where \(\eta_i \in \mathcal{H}_{\mathcal{S}}\) follows from the association between the kernel integral operator and its associated RKHS \citep{cucker2002mathematical}. Accordingly, by the reproducing property of the tensor-product RKHS, the above criterion admits the inner-product representation
\begin{equation}\label{eq:empirical-functional-loss}
    \ell_n(\beta) = \frac{1}{2n} \sum_{i=1}^{n} \int_{\mathcal{T}} \left\{ Y_i(t) - \langle\beta,\Psi_{i,t}\rangle_{\mathcal{H}} \right\}^{2}\,dt,
    \qquad
    \beta\in\mathcal{H}.
\end{equation}
However, directly minimizing \eqref{eq:empirical-functional-loss} requires access to the complete sample and is therefore infeasible for streaming data. 
Instead, we propose a one-pass estimator by applying SGD to the instantaneous loss \(L_i(\beta) = \int_{\mathcal{T}}\{Y_i(t)-\langle\beta,\Psi_{i,t}\rangle_{\mathcal{H}}\}^2\,dt/2\). Define
\begin{equation}\label{eq:R_C_Definition}
    R_i = \int_{\mathcal{T}}Y_i(t)\Psi_{i,t}\,dt \in \mathcal{H} \quad \text{and} \quad C_i = \int_{\mathcal{T}}(\Psi_{i,t} \otimes_{\mathcal{H}} \Psi_{i,t})~dt
\end{equation}
respectively, where $C_i$ is the individual covariance operator and \(u\otimes_{\mathcal{H}}v\) denotes the rank-one operator defined as \((u\otimes_{\mathcal{H}}v)f=\langle v,f\rangle_{\mathcal{H}}u\) for \(u,v,f\in\mathcal{H}\). Then the Fréchet gradient of $L_i(\beta)$ in \(\mathcal{H}\) is explicitly given by \(\nabla L_i(\beta) = C_i\beta - R_i\). Given an initial value \(\widehat\beta_0\in\mathcal{H}\) and a positive step size \(\gamma_i\), the functional SGD estimator is updated 
upon the arrival of \(D_i=(X_i,Y_i)\) 
by
\begin{equation}\label{eq:functional-sgd-update}
    \widehat\beta_i = \widehat\beta_{i-1} - \gamma_i \nabla L_i(\widehat\beta_{i-1}) = 
    \bigl(I_{\mathcal{H}}-\gamma_i C_i\bigr) \widehat\beta_{i-1} + \gamma_i R_i,
    \qquad
    i=1,\ldots,n,
\end{equation}
where \(I_{\mathcal{H}}:\mathcal{H}\to\mathcal{H}\) denotes the identity operator. By mitigating stochastic noise in individual iterates \citep{bach2013non}, averaging enables stochastic gradient methods to achieve optimal prediction rates in nonparametric RKHS regression with an appropriate step size and regularity conditions \citep{dieuleveut2016nonparametric}.
Motivated by this, we define the Polyak--Ruppert averaged estimator \(\bar\beta_n = n^{-1}\sum_{i=1}^{n}\widehat\beta_i\) after \(n\) observations \citep{ruppert1988efficient,polyak1992acceleration}. 
In light of \eqref{eq:functional-sgd-update}, $\bar{\beta}_i$ can be updated recursively as 
\begin{equation}\label{eq:beta_bar_recursion}
    \bar{\beta}_i = \frac{i-1}{i}\bar{\beta}_{i-1} + \frac{1}{i}\widehat{\beta}_i, \qquad
    i=1,\ldots,n,
\end{equation}
and $    \bar{\beta}_0=\widehat{\beta}_0$. Notably, both \(\widehat\beta_i\) and \(\bar\beta_i\) are updated using only the incoming \(D_i\) and the state at Step \(i-1\). Once updated, \(D_i\) is 
discarded, preserving a strictly one-pass procedure.

\subsection{Online multiplier bootstrap inference}\label{subsec:online-bootstrap}

In this subsection, we focus on inference for the bivariate slope surface \(\beta_0\). 
In particular, we construct pointwise confidence intervals for $\beta_0$ based on the averaged estimator \(\bar\beta_n\) computed via \eqref{eq:beta_bar_recursion}.
For a fixed point \(z=(t,s)\in\mathcal{T} \times \mathcal{S}\), we aim to approximate the sampling distribution of the pointwise estimation error
\(\bar\beta_n(z)-\beta_0(z)\). To this end, an online multiplier bootstrap is developed based on a randomly perturbed functional SGD recursion.

Let \(\{\omega_i\}_{i\geq 1}\) be a sequence of i.i.d.\ nonnegative multiplier variables, independent of \(\mathcal{D}_n\), satisfying \(\mathbb{E}(\omega_i) = \operatorname{Var}(\omega_i) = 1\) and \(0\leq \omega_i\leq \omega_{\max}<\infty\) almost surely for some constant \(\omega_{\max}\). Starting from \(\widehat\beta_0^*=\widehat\beta_0\), the perturbed functional SGD iterate is updated according to
\begin{equation}\label{eq:bootstrap-update}
    \widehat\beta_i^*
    = \widehat\beta_{i-1}^* - \gamma_i\omega_i \bigl(C_i\widehat\beta_{i-1}^*-R_i\bigr)
    = \bigl(I_{\mathcal{H}}-\gamma_i\omega_i C_i\bigr) \widehat\beta_{i-1}^* + \gamma_i\omega_i R_i,
    \qquad
    i=1,\ldots,n,
\end{equation}
yielding the perturbed averaged estimator \(\bar\beta_n^* = n^{-1}\sum_{i=1}^n\widehat\beta_i^*\), which can also be updated by
\begin{equation}\label{eq:bootstrap-beta-bar-recursion}
    \bar\beta_i^*
    = \frac{i-1}{i}\bar\beta_{i-1}^*
    + \frac{1}{i}\widehat\beta_i^*,
    \qquad
    \bar\beta_0^*=\widehat\beta_0^*.
\end{equation}
This extends the online perturbation paradigm from finite-dimensional parameters \citep{fang2018online} and scalar-on-function regression \citep{xie2026scalable} to bivariate surface inference.

In practice, we generate \(B\) independent multiplier sequences \(\{\omega_i^b\}_{i=1}^n\) (\(b=1,\ldots,B\)) and denote by \(\bar\beta_n^{*,b}\) the resulting perturbed averages obtained from \eqref{eq:bootstrap-update}. For a fixed \(z=(t,s)\in\mathcal{T}\times\mathcal{S}\), let \(\Delta_n^{*,b}(z)=\bar\beta_n^{*,b}(z)-\bar\beta_n(z)\). Conditional on \(\mathcal{D}_n\), the empirical distribution of \(\{\Delta_n^{*,b}(z)\}_{b=1}^B\) is obtained, with \(\widehat q_\tau(z)\) being its \(\tau\)-quantile. Accordingly, a bootstrap-quantile (B-Q) pointwise confidence interval for \(\beta_0(z)\) with a nominal coverage \(1-\delta\) is
\[
    \mathrm{CI}_{1-\delta}^{\mathrm{B\text{-}Q}}(z)
=
\left[
\bar\beta_n(z)-\widehat q_{1-\delta/2}(z),
\,
\bar\beta_n(z)-\widehat q_{\delta/2}(z)
\right].
\]
The theoretical validity of this bootstrap approximation, including the required bias conditions, is established in Section~\ref{sec:theory}. Note that at each step \(i\), \(R_i\) and \(C_i\) are computed and shared across all recursions before \(D_i\) is discarded. Therefore, the online multiplier bootstrap procedure can be implemented in a parallel manner to enhance computational efficiency. The joint updates can be seen in Algorithm~\ref{alg:online-bootstrap-fof-sgd}.

\begin{algorithm}[!ht]
\caption{Online function-on-function SGD with multiplier bootstrap}
\label{alg:online-bootstrap-fof-sgd}

\textbf{Input:}
Data stream \(\{D_i=(X_i,Y_i)\}_{i=1}^n\);
step sizes \(\{\gamma_i\}_{i=1}^n\);
number of bootstrap replications \(B\);
initial value \(\widehat\beta_0\).

\begin{algorithmic}[1]

\State Set \(\bar\beta_0\gets\widehat\beta_0\) and \(\widehat\beta_0^{*,b} =\bar\beta_0^{*,b} \gets\widehat\beta_0\) for \(b=1,\ldots,B\).

\For{\(i=1,\ldots,n\)}

    \State 
    Observe \(D_i\) and compute \(R_i\) and \(C_i\) using \eqref{eq:R_C_Definition}.

    \State Update \(\widehat\beta_i\) and \(\bar\beta_i\) using \eqref{eq:functional-sgd-update} and \eqref{eq:beta_bar_recursion}, respectively.

\For{\(b=1,\ldots,B\)}

    \State Generate \(\omega_i^b\), then update \(\widehat\beta_i^{*,b}\) via \eqref{eq:bootstrap-update} and \(\bar\beta_i^{*,b}\) via \eqref{eq:bootstrap-beta-bar-recursion}.

\EndFor

\State Discard \(D_i\).

\EndFor

\end{algorithmic}

\textbf{Output:} \(\bar\beta_n\) and \(\{\bar\beta_n^{*,b}\}_{b=1}^B\).

\end{algorithm}

\subsection{Practical online selection for step size}\label{subsec:adaptive-step}

The empirical performance of stochastic gradient methods can be sensitive to the choice of step size \citep{dieuleveut2016nonparametric, bottou2018optimization}. To avoid manual tuning, we adapt the wRV procedure proposed by \citet{ZhangLei2025} to the proposed online function-on-function regression setting, which selects among a finite collection of fixed-step candidates via rolling prediction losses. 
Defining \(\Theta=\{c_k:k=1,\ldots,K\}\) with \(c_k>0\), Candidate \(k\) uses the fixed step size \(\gamma_{n,k}=c_kn^{-\zeta}\) for a constant \(\zeta>0\). Starting from \(\widehat\beta_0^{(k)}=\widehat\beta_0\), the \(k\)-th stochastic gradient trajectory is updated by
\begin{equation}\label{eq:wrv-fixed-step-update}
    \widehat\beta_i^{(k)} = \bigl(I_{\mathcal{H}}-\gamma_{n,k}C_i\bigr)
\widehat\beta_{i-1}^{(k)} + \gamma_{n,k}R_i,
\qquad
i=1,\ldots,n,
\end{equation}
with the corresponding averaged estimator \(\bar\beta_i^{(k)}\) updated recursively according to \eqref{eq:beta_bar_recursion}, and all the candidates are compared using a predict-then-update scheme. Specifically for \(i\geq2\), Candidate \(k\) first produces the prediction and incurs the loss
\begin{equation}\label{eq:wrv-preupdate-prediction}
    \widehat Y_{i\mid i-1}^{(k)}(t)
    = \int_{\mathcal{S}}X_i(s)\bar\beta_{i-1}^{(k)}(t,s)\,ds,
    \qquad
    \ell_{i,k}
    = \|Y_i-\widehat Y_{i\mid i-1}^{(k)}\|_{L^2(\mathcal{T})}^2,
\end{equation}
where \(t\in\mathcal{T}\).
For a fixed \(\xi\geq 0\), let \(w_i=(i-1)^\xi\) (\(i=2,\ldots,n\)) and define
\begin{equation}
\operatorname{wRV}_{n,k} = \frac{\sum_{i=2}^{n}w_i\ell_{i,k}}{\sum_{i=2}^{n}w_i},
\qquad
k=1,\ldots,K.
\label{eq:wrv-criterion}
\end{equation}
After \(n\) observations, we select \(\widehat k_n=\arg\min_{1\leq k\leq K}\operatorname{wRV}_{n,k}\) and report \(\bar\beta_n^{\mathrm{wRV}}=\bar\beta_n^{(\widehat k_n)}\). The complete wRV procedure is provided in Section~S.1 of the Supplementary Material.

\section{Asymptotic properties}\label{sec:theory}

We develop the asymptotic theory for the proposed online estimation and inference procedures in this section. Section~\ref{subsec:convergence-rates} establishes convergence rates of prediction and estimation for the proposed estimator, while Section~\ref{subsec:bootstrap-inference} establishes the validity of the online multiplier bootstrap. Throughout this section, we take \(\widehat\beta_0=\widehat\beta_0^*=0\).

\subsection{Convergence rates for prediction and estimation}\label{subsec:convergence-rates}

To evaluate the statistical performance of the proposed online estimator, we first consider its prediction accuracy on a future observation \(D_{n+1}=(X_{n+1},Y_{n+1})\). Let \(\mathcal{F}_n=\sigma(D_1,\ldots,D_n)\) denote the \(\sigma\)-field generated by the first \(n\) pairs of observations. Suppose \(D_{n+1}\) has the same distribution as \(D_1\) and is independent of \(\mathcal{F}_n\). Let \(\mathbb{E}_{n+1}(\cdot)\) denote the expectation with respect to \(D_{n+1}\), conditional on \(\mathcal{F}_n\). For any \(\mathcal{F}_n\)-measurable estimator \(\widetilde\beta_n\in\mathcal{H}\), define the conditional excess prediction risk by
\[
    \mathcal{R}_n^{\mathrm{pred}}(\widetilde\beta_n) = \int_{\mathcal{T}} \mathbb{E}_{n+1} \Bigl[ \bigl\{ Y_{n+1}(t) - \langle \widetilde\beta_n, \Psi_{n+1,t} \rangle_{\mathcal{H}} \bigr\}^{2} 
    -
    \bigl\{ Y_{n+1}(t) - \langle \beta_0, \Psi_{n+1,t} \rangle_{\mathcal{H}} \bigr\}^{2} \Bigr]\,dt,
\]
where \(\Psi_{n+1,t}\) is defined as in \eqref{eq:eta_Psi_definition}. Set \(V_i=\int_{\mathcal{T}}\|\Psi_{i,t}\|_{\mathcal{H}}^2\,dt\) and assume \(\mathbb{E}(V_i)<\infty\). Recall the individual covariance operator \(C_i\) given in \eqref{eq:R_C_Definition}.
Since \(C_i\) is positive and trace-class with \(\operatorname{tr}_{\mathcal{H}}(C_i)=V_i\), the population covariance operator \(\Sigma_{\mathcal{H}} = \mathbb{E}(C_i)\) is positive, self-adjoint, trace-class, and therefore compact. Let \(\{(\nu_j,\phi_j)\}_{j\geq1}\) be its nonzero eigenpairs, with \(\nu_1\geq\nu_2\geq\cdots>0\) and \(\Sigma_{\mathcal{H}}\phi_j=\nu_j\phi_j\), where \(\{\phi_j\}_{j\geq1}\) is orthonormal in \(\mathcal{H}\). For \(f\in\mathcal{H}\), define \(\|f\|_{\Sigma_{\mathcal{H}}}^{2} = \langle f,\Sigma_{\mathcal{H}}f\rangle_{\mathcal{H}}\). The seminorm \(\|\cdot\|_{\Sigma_{\mathcal{H}}}\), induced by the population covariance structure, provides the natural prediction metric used in \citet{xie2026scalable}.

In the present function-on-function setting, the following proposition shows that the conditional excess prediction risk is exactly the squared \(\Sigma_{\mathcal{H}}\)-seminorm of the estimation error, equivalent to the covariance-weighted bilinear form representation in \citet{dette2024statistical}. 

\begin{proposition}\label{prop:proposition1}
For any \(\mathcal{F}_n\)-measurable estimator \(\widetilde\beta_n\in\mathcal{H}\), \(\mathcal{R}_n^{\mathrm{pred}}(\widetilde\beta_n) = \|\widetilde\beta_n-\beta_0\|_{\Sigma_{\mathcal{H}}}^{2}.\)
\end{proposition}

Next we establish the prediction accuracy of the Polyak--Ruppert averaged estimator under the following regularity conditions.

\begin{assumption}\label{ass:assumption1}
The nonzero eigenvalues \(\{\nu_j\}_{j\geq1}\) of \(\Sigma_{\mathcal{H}}\), arranged in a decreasing order, satisfy \(s_1j^{-\alpha} \leq \nu_j \leq s_2j^{-\alpha}\) for all \(j\geq1\), where \(0<s_1<s_2\) and \(\alpha>1\) are constants. 
\end{assumption}

\begin{assumption}\label{ass:assumption2}
The true slope surface belongs to the closed linear span of \(\{\phi_j:j\geq1\}\) and admits the expansion \(\beta_0 = \sum_{j=1}^{\infty}a_j\phi_j\) in \(\mathcal{H}\), where \(a_j = \langle\beta_0,\phi_j\rangle_{\mathcal{H}}.\) There exist constants \(C_a>0\) and \(\tau>\alpha/2+1\) such that \(|a_j| \leq C_a j^{-\tau}\) for all \(j\geq1\).
\end{assumption}

\begin{assumption}\label{ass:assumption3}
There exists a constant \(R>0\) such that \(\mathbb{E}(V_iC_i) \preceq R^2\Sigma_{\mathcal{H}},\) where \(\preceq\) denotes the canonical partial order for a self-adjoint operator on \(\mathcal{H}\).
\end{assumption}

\begin{remark}\label{rem:assumptions1-3}
Assumption~\ref{ass:assumption1} imposes polynomial decay on the nonzero eigenvalues of \(\Sigma_{\mathcal{H}}\), as in \citet{xie2026scalable}. Similar polynomial spectral conditions can be found in \citet{sun2018optimal}. Assumption~\ref{ass:assumption2} imposes polynomial decay on the slope coefficients relative to the population eigensystem, parallel to the condition in \citet{HallHorowitz2007}. The restriction \(\tau>\alpha/2+1\) is necessary to control the bias, while the expansion over the nonzero eigenspaces ensures identifiability. Indeed, Assumptions~\ref{ass:assumption1} and~\ref{ass:assumption2} imply \(\sum_{j=1}^{\infty} a_j^2/\nu_j < \infty\). Assumption~\ref{ass:assumption3} is the functional-response analogue of the covariance-adapted fourth-moment condition in \citet{xie2026scalable} and shares the operator form of the least-squares moment condition in \citet{bach2013non}. It holds, in particular, if \(V_i\leq R^2\) almost surely, since \(V_iC_i\preceq R^2C_i\) and \(\mathbb{E}(C_i)=\Sigma_{\mathcal{H}}\). In fact, if $X_i$ is a uniformly bounded process, there must exist some $R$ such that $V_i \le R^2$ almost surely.
\end{remark}

\begin{theorem}[Prediction accuracy]\label{thm:theorem1}
Suppose Assumptions~\ref{ass:assumption1}--\ref{ass:assumption3} hold. For any constant step size \(\gamma>0\) satisfying \(\gamma R^2<1\), \(\gamma\nu_1\leq1\), and \(n\gamma\geq1\), it holds that
\[
    \mathbb{E} \|\bar\beta_n-\beta_0\|_{\Sigma_{\mathcal{H}}}^{2} = \mathcal{O} \left( C_a^2 \left\{ (n\gamma)^{-2}
    + n^{-1}\right\}
    +
    \sigma^2 \frac{\gamma^{1/\alpha}} {n^{1-1/\alpha}} \right).
\]
\end{theorem}

\begin{corollary}[Prediction accuracy with an optimal constant step size]\label{cor:corollary1} 
Suppose the conditions in Theorem~\ref{thm:theorem1} hold. If \(\gamma = \gamma_n \asymp n^{-(\alpha+1)/(2\alpha+1)},\) then
\[
    \mathbb{E} \|\bar\beta_n-\beta_0\|_{\Sigma_{\mathcal{H}}}^{2} = \mathcal{O} \left( n^{-2\alpha/(2\alpha+1)} \right).
\]
\end{corollary}

\begin{remark}
    Theorem~\ref{thm:theorem1} provides a prediction bound for the proposed averaged estimator. Consequently, the leading bias and variance terms are \((n\gamma)^{-2}\) and \(\gamma^{1/\alpha}n^{-1+1/\alpha}\), respectively. Balancing these terms yields \(\gamma\asymp n^{-(\alpha+1)/(2\alpha+1)}\) and the optimal prediction rate \(n^{-2\alpha/(2\alpha+1)}\), which takes the same functional form as the offline minimax prediction rates established by \citet{sun2018optimal} and \citet{dette2024statistical} under their respective spectral conditions.
\end{remark}
 
Unlike the prediction error, bounding the \(L^2(\mathcal{T}\times\mathcal{S})\) estimation error requires controlling the fluctuations of the individual covariance operator \(C_i\). To this end, we recall the predictor representation \(\eta_i\) given in \eqref{eq:eta_Psi_definition} and define its second-moment operator \(\Gamma_{\eta} = \mathbb{E}\bigl(\eta_i \otimes_{\mathcal{H}_{\mathcal{S}}} \eta_i\bigr)\), where \((u\otimes_{\mathcal{H}_{\mathcal{S}}}v)f=\langle v,f\rangle_{\mathcal{H}_{\mathcal{S}}}u\) for \(u,v,f\in\mathcal{H}_{\mathcal{S}}\).

\begin{assumption}\label{ass:assumption4}
There exists a constant \(\kappa_X<\infty\) such that, for every \(u\in\mathcal{H}_{\mathcal{S}}\), \(\mathbb{E} \bigl| \langle\eta_i,u\rangle_{\mathcal{H}_{\mathcal{S}}} \bigr|^4 \leq \kappa_X \bigl\{ \langle \Gamma_{\eta}u, u \rangle_{\mathcal{H}_{\mathcal{S}}} \bigr\}^{2}\), where \(\eta_i = \int_{\mathcal{S}} X_i(s)K_{\mathcal{S}}(s,\cdot)\,ds \in \mathcal{H}_{\mathcal{S}}\).
\end{assumption}

\begin{remark}\label{rem:assumption4}
Assumption~\ref{ass:assumption4} imposes a uniform kurtosis bound on the linear projections of the predictor representation \(\eta_i\). Indeed, by the reproducing property, \(\langle\eta_i,u\rangle_{\mathcal{H}_{\mathcal{S}}} = \int_{\mathcal{S}}X_i(s)u(s)\,ds\) for all \(u\in\mathcal{H}_{\mathcal{S}}\), so this assumption is the RKHS-representation counterpart of Condition~(41) in \citet{yuan2010reproducing}. It holds with \(\kappa_X=3\) for centered Gaussian predictor representations and, more generally, for representations with independent standardized scores having uniformly bounded fourth moments.
\end{remark}

\begin{theorem}[Estimation accuracy]\label{thm:theorem2}
Suppose Assumptions~\ref{ass:assumption1}--\ref{ass:assumption4} hold, and let \(\widetilde\tau = \min\{\tau, \alpha+1/2\}\). For any constant step size \(\gamma>0\) satisfying \(n\gamma\to\infty\) and \(\gamma=o(n^{-1/(\alpha+1)})\), the averaged estimator satisfies
\[
    \mathbb{E} \|\bar\beta_n-\beta_0\|_{L^2(\mathcal{T}\times\mathcal{S})}^2 = \mathcal{O}
    \left( C_a^2 (n\gamma)^{-(2\widetilde\tau-1)/\alpha}\{\log(e n\gamma)\}^{\mathbf{1}\{\tau=\alpha+1/2\}} + (C_a^2+\sigma^2) \gamma(n\gamma)^{1/\alpha} \right).
\]
\end{theorem}

\begin{corollary}[Estimation accuracy with an optimal constant step size]\label{cor:corollary2}
Under the conditions in Theorem~\ref{thm:theorem2}, choosing
$\gamma = \gamma_n \asymp n^{-2\widetilde\tau/(2\widetilde\tau+\alpha)} (\log n)^{\{\alpha/(3\alpha+1)\}\mathbf{1}\{\tau=\alpha+1/2\}}
$ yields
\[
    \mathbb{E} \|\bar\beta_n-\beta_0\|_{L^2(\mathcal{T}\times\mathcal{S})}^2 = \mathcal{O} \left( n^{-(2\widetilde\tau-1)/(2\widetilde\tau+\alpha)} (\log n)^{\{(\alpha+1)/(3\alpha+1)\}\mathbf{1}\{\tau=\alpha+1/2\}} \right).
\]
\end{corollary}

\begin{remark}
For \(\alpha/2+1<\tau<\alpha+1/2\), the rate \(n^{-(2\tau-1)/(2\tau+\alpha)}\) matches the minimax integrated squared error rate for function-on-function regression \citep{imaizumi2018pca} and its scalar-response counterpart \citep{HallHorowitz2007}. At the boundary \(\tau=\alpha+1/2\), an additional logarithmic factor appears. When \(\tau>\alpha+1/2\), the uniform Polyak--Ruppert average exhibits saturation, i.e., further coefficient smoothness does not improve the leading \(L^2\)-bias order beyond \((n\gamma)^{-2}\). This behavior is analogous to the saturation of uniformly averaged SGD in \citet{mucke2019beating}.
\end{remark}

\subsection{Pointwise bootstrap validity}\label{subsec:bootstrap-inference}

Next, we establish the pointwise validity of the online bootstrap procedure for the averaged estimator \(\bar\beta_n(z_0)\), where \(z_0=(t_0,s_0)\in\mathcal{T}\times\mathcal{S}\) is fixed. Recall that \(\{(\nu_j,\phi_j)\}_{j\geq1}\) denotes the nonzero eigenpairs of \(\Sigma_{\mathcal{H}}\), with \(\{\phi_j\}_{j\geq1}\) orthonormal in \(\mathcal{H}\). For \(i,j\geq1\), define the spectral score coordinates
\(U_{ij}=\int_{\mathcal{T}}\varepsilon_i(t)\langle\Psi_{i,t},\phi_j\rangle_{\mathcal{H}}\,dt\).
Since the data pairs are i.i.d.\ and \(\mathbb{E}(\varepsilon_i\mid X_i)=0\), the sequence \(\{U_{ij}\}_{i\geq1}\) is i.i.d.\ and centered for each \(j\). Thus, \(\rho_j:=\mathbb{E}(U_{ij}^2)\) does not depend on \(i\). We further impose the following conditions on these coordinates.

\begin{assumption}\label{ass:assumption5}
There exist constants \(C_\rho>0\), \(\delta>0\), and \(m_1>1\) such that \(\rho_j\asymp j^{-m_1}\). In addition, for every \(i\geq1\) and every deterministic sequence \(\{b_j\}_{j\geq1}\) satisfying \(\sum_{j\geq1}b_j^2\rho_j<\infty\), we have \(\sum_{j,r\geq1} | b_jb_r\,\mathbb{E}(U_{ij}U_{ir}) | \leq C_\rho\sum_{j\geq1}b_j^2\rho_j\) and \(\mathbb{E}\bigl| \sum_{j\geq1}b_jU_{ij} \bigr|^{2+\delta} \leq C_\rho \bigl( \mathbb{E}\bigl| \sum_{j\geq1}b_jU_{ij} \bigr|^2 \bigr)^{1+\delta/2}\).
\end{assumption}

For each sample size \(n\), let \(\gamma_i=\gamma_n\asymp n^{-\zeta}\) (\(1\leq i\leq n\)) for some \(0<\zeta<1\). For \(1\leq k\leq n\), define \(\ell_{n,k}(z_0) = n^{-1} \sum_{j=1}^{\infty} \nu_j^{-1} \{1-(1-\gamma_n\nu_j)^{n+1-k}\} U_{kj}\phi_j(z_0)\) and \(v_n^2(z_0) = \sum_{k=1}^{n} \mathbb{E}\{\ell_{n,k}^2(z_0)\}\).

\begin{assumption}\label{ass:assumption6}
For the fixed point \(z_0\), \(0<v_n^2(z_0)<\infty\) for all sufficiently large \(n\), and \(v_n^{-2}(z_0) \allowbreak \max_{1\leq k\leq n} \mathbb{E}\{\ell_{n,k}^2(z_0)\} \allowbreak \to 0\) as \(n\to\infty\).
\end{assumption}

\begin{remark}\label{rem:assumptions5-6}
Assumption~\ref{ass:assumption5} controls the decay, dependence, and higher-order moments of the spectral score coordinates. The condition \(m_1>1\) ensures \(\sum_{j\geq1}\rho_j<\infty\). If the coordinates are uncorrelated, i.e., \(\mathbb{E}(U_{ij}U_{ir})=\rho_j\mathbf{1}\{j=r\}\), the covariance summability bound holds with \(C_\rho=1\). This occurs, for instance, when 
the covariance operator of \(\int_{\mathcal T} \varepsilon_i(t)\Psi_{i,t}\,dt\)
and \(\Sigma_{\mathcal{H}}\) share the eigenfunctions \(\{\phi_j\}_{j\geq1}\) \citep{xie2026scalable}. Assumption~\ref{ass:assumption5} is also satisfied by independent centered Gaussian coordinates with \(\rho_j\asymp j^{-m_1}\). Assumption~\ref{ass:assumption6} controls the variance of the leading pointwise stochastic term. Since \(\ell_{n,k}(z_0)\) are independent and centered, \(v_n^2(z_0)=\operatorname{Var}\bigl\{\sum_{k=1}^{n}\ell_{n,k}(z_0)\bigr\}\). Its positivity ensures a well-defined pointwise normalization for large \(n\), while the maximal-variance condition prevents any observation from dominating the total variance. Combined with the \(2+\delta\) moment bound in Assumption~\ref{ass:assumption5}, this yields the Lyapunov condition for the pointwise central limit theorem \citep{billingsley2017probability}.
\end{remark}

\begin{theorem}\label{thm:theorem3}
Suppose Assumptions~\ref{ass:assumption1}--\ref{ass:assumption6} hold. Assume that \(\gamma_n\omega_{\max}^2R^2<1\), \(\gamma_n\nu_1\leq1\), and \(n^{-1}+\gamma_n^2=o\bigl(v_n^2(z_0)\bigr)\) for all sufficiently large \(n\). Then, for any fixed \(z_0=(t_0,s_0)\in\mathcal{T}\times\mathcal{S}\),
\[
\sup_{\kappa\in\mathbb{R}} \left| \mathbb{P}^* \left( \frac{\bar\beta_n^*(z_0)-\bar\beta_n(z_0)} {v_n(z_0)} \leq\kappa \right) - \mathbb{P} \left( \frac{\bar\beta_n(z_0)-\beta_0(z_0) -\operatorname{Bias}_n(z_0)} {v_n(z_0)} \leq\kappa \right) \right| \xrightarrow{\mathbb{P}}0,
\]
where \(\mathbb{P}^*(\cdot)=\mathbb{P}(\cdot\mid\mathcal{F}_n)\) and \(\operatorname{Bias}_n(z_0) = - \bigl( n^{-1} \sum_{k=1}^{n} \bigl(I_{\mathcal{H}}-\gamma_n\Sigma_{\mathcal{H}}\bigr)^k \beta_0 \bigr)(z_0)\). If \(\operatorname{Bias}_n(z_0)=o\bigl(v_n(z_0)\bigr)\), the same conclusion holds with \(\operatorname{Bias}_n(z_0)\) being omitted.
\end{theorem}

\begin{corollary}\label{cor:corollary3}
Suppose Assumptions~\ref{ass:assumption1}--\ref{ass:assumption5} hold with \(\mathbb{E}(U_{ij}U_{ir})=\rho_j\mathbf{1}\{j=r\}\) for all \(i,j,r\geq1\). For a fixed \(z_0=(t_0,s_0)\in\mathcal{T}\times\mathcal{S}\), assume \(\sum_{j=m}^{2m}\phi_j^2(z_0)\asymp m^{1-2\chi}\) for some \(\chi>1/2\). Assume \(m_\star:=(2\alpha-m_1-2\chi+1)/\alpha>0\) and \(q_0=\min\{1,(\tau+\chi-1)/\alpha\}\). If \(\gamma_n\asymp n^{-\zeta}\) and \(\max\{0,(1-m_\star)/(2-m_\star)\} < \zeta < 1\), \(v_n^2(z_0)\asymp n^{-1}(n\gamma_n)^{m_\star}\), Assumption~\ref{ass:assumption6} holds, and \(n^{-1}+\gamma_n^2=o\bigl(v_n^2(z_0)\bigr)\). Moreover,
\[
\sup_{\kappa\in\mathbb{R}} \left| \mathbb{P}^* \left( \frac{\bar\beta_n^*(z_0)-\bar\beta_n(z_0)}{n^{-1/2}(n\gamma_n)^{m_\star/2}} \leq\kappa \right) - \mathbb{P} \left( \frac{\bar\beta_n(z_0)-\beta_0(z_0)-\operatorname{Bias}_n(z_0)}{n^{-1/2}(n\gamma_n)^{m_\star/2}} \leq\kappa \right) \right| \xrightarrow{\mathbb{P}}0.
\]
If, in addition, \(\zeta<1-(2q_0+m_\star)^{-1}\), then \(\operatorname{Bias}_n(z_0)=o\bigl(v_n(z_0)\bigr)\), and the same conclusion holds with \(\operatorname{Bias}_n(z_0)\) being omitted.
\end{corollary}

\begin{remark} \label{rem:bootstrap-validity}
Theorem~\ref{thm:theorem3} extends the pointwise online bootstrap framework in \citet{xie2026scalable} to function-on-function regression using the recursive multiplier construction of \citet{fang2018online}. The quantity \(v_n(z_0)\) is the standard deviation of the leading linearized stochastic term, while the condition \(n^{-1}+\gamma_n^2=o\bigl(v_n^2(z_0)\bigr)\) ensures that the random-operator remainder is asymptotically negligible. Corollary~\ref{cor:corollary3} gives \(v_n^2(z_0)\asymp n^{-1}(n\gamma_n)^{m_\star}\), together with a step size condition for bias negligibility. When the additional bias-negligibility condition in Corollary~\ref{cor:corollary3} holds, conditional bootstrap quantiles yield asymptotically valid pointwise inference for \(\beta_0(z_0)\) without requiring explicit estimation of \(v_n(z_0)\).
\end{remark}

\section{Simulation studies}\label{sec:simulation}
In this section, we evaluate the finite sample performance of the proposed method through extensive simulation studies. 
We first examine the prediction and estimation accuracy of the proposed estimator under three simulation scenarios and investigate its sensitivity to the step size. 
Then we assess the wRV procedure and compare the resulting online estimator with the corresponding offline estimator of \cite{sun2018optimal}. Finally, we study the empirical coverage probabilities and interval lengths of the pointwise confidence intervals constructed by the proposed online bootstrap procedure. Throughout the simulation  study, we consider the second-order Sobolev space for $\mathcal H_{\mathcal T}$ and $\mathcal H_{\mathcal S}$, i.e., $m = 2$, and choose $\hat{\beta}_0 = \hat{\beta}_0^* = 0$ in our numerical implementations.

\subsection{Estimation of \texorpdfstring{\(\beta_0\)}{beta0}}
\label{subsec:estimation}

To start with, we generate data from model \eqref{eq:base_model} with different structures in \(\beta_0(t, s)\), with \(\mathcal{T}=\mathcal{S}=[0,1]\), respectively. Specifically, the predictor process \(X_i(s)\) is generated as \(\sum_{k=1}^{50}(-1)^{k+1}k^{-1} \allowbreak Z_{ik}\vartheta_1(s,k)\), where \(Z_{ik}\) are independent \(\mathrm{Uniform}[-\sqrt{3}, \allowbreak \sqrt{3}]\) variables, with basis functions \(\vartheta_1(s,1)=1\) and \(\vartheta_1(s,k) \allowbreak = \sqrt{2}\cos\{(k-1)\pi s\}\) (\(k\geq 2\)), while the slope surface is specified as
\begin{itemize}
    \item \refstepcounter{scenario}\label{scen:1} \textbf{Scenario \arabic{scenario}.} 
    \(\beta_0(t,s)=\exp\{-(t+s)\}\);
    
    \item \refstepcounter{scenario}\label{scen:2} \textbf{Scenario \arabic{scenario}.} 
    \(\beta_0(t,s)=\sum_{k=1}^{3}(-1)^{k+1}k^{-3} \vartheta_1(t,k)\vartheta_1(s,k)\);
    
    \item \refstepcounter{scenario}\label{scen:3} \textbf{Scenario \arabic{scenario}.} 
    \(\beta_0(t,s)=t^2+s^2+ts\). 
    
\end{itemize}

Scenario~\ref{scen:1} follows from that in \citet{sun2018optimal} and corresponds to the second setting in \citet{dette2024statistical}. Scenario~\ref{scen:2} is similar to those in \citet{yuan2010reproducing,sun2018optimal} and \citet{dette2024statistical}, while the third is an additional polynomial setting. In all settings, \(X_i\) and \(Y_i\) are evaluated at \(101\) equally spaced points on \([0,1]\), with i.i.d.\ \(N(0,0.5^2)\) response-grid errors that are independent of \(X_i\). 
We evaluate the performance of the proposed method across sequential sample sizes \(n \in \{500, 1000, 5000, 10000\}\) arriving in data streams of length \(10000\). The step size is fixed as \(\gamma_n = n^{-\zeta}\), with \(\zeta\) being selected from \(\{0.0001, 0.001, 0.01, 0.05, 0.1, 0.2\}\), so as to investigate the impact of the step size on the estimation of $\beta_0(t, s)$. Performance is quantified by the squared \(L^2\) estimation error \(\operatorname{EE}(\bar\beta_n) = \int_{\mathcal{T}}\int_{\mathcal{S}}\{\bar\beta_n(t,s)-\beta_0(t,s)\}^2\,ds\,dt\), and the prediction error \(\operatorname{PE}(\bar\beta_n) = n_{\mathrm{test}}^{-1}\sum_{i=1}^{n_{\mathrm{test}}}\int_{\mathcal{T}}\bigl[\int_{\mathcal{S}}X_i^{\mathrm{test}}(s)\{\bar\beta_n(t,s)-\beta_0(t,s)\}\,ds\bigr]^2\,dt\) computed on an independent test sample of \(n_{\mathrm{test}}=1000\) predictor trajectories. All reported metrics are averaged over \(M=200\) independent Monte Carlo replications. 

\begin{figure}[!ht]
    \centering
    \includegraphics[width=\linewidth]{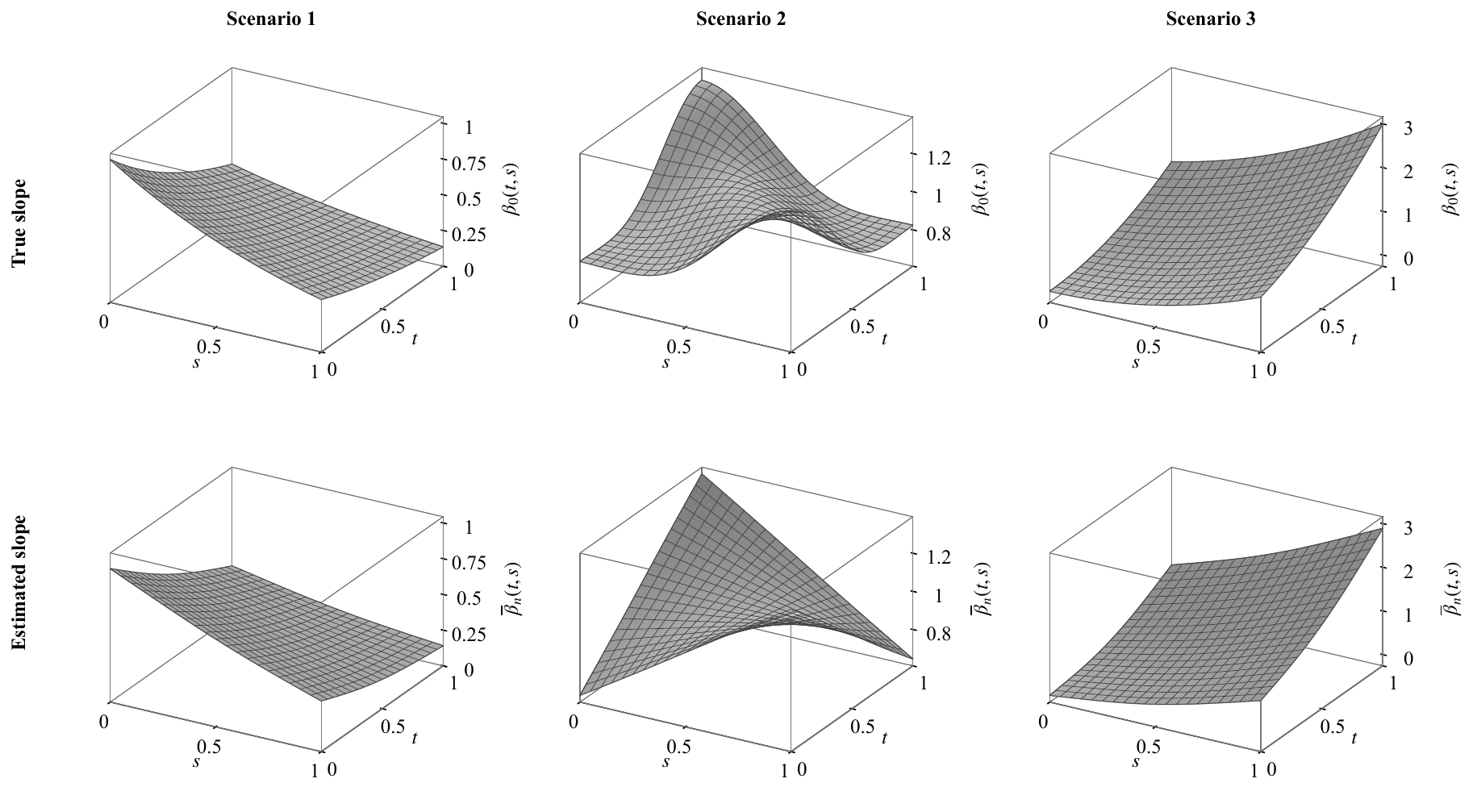}
    \caption{
    The true slope surfaces \(\beta_0\) and the corresponding averaged online estimates under Scenarios~\ref{scen:1}--\ref{scen:3} at \(n=10000\).
    }
    \label{fig:beta_surface_comparison}
\end{figure}

Table~\ref{tab:simulation_errors_zero} summarizes the Monte Carlo averages of the prediction errors under Scenarios~\ref{scen:1}--\ref{scen:3}, and their standard errors. Overall, the prediction errors decrease consistently as the sample size grows, regardless of the choice of \(\zeta\). In Scenario~\ref{scen:1}, \(\zeta=0.01\) generally performs best, although \(\zeta=0.05\) gives a slightly smaller prediction error at \(n=500\). In Scenario~\ref{scen:2}, the prediction errors decrease more slowly with the sample size, and \(\zeta=0.0001\) and \(0.001\) yield comparable and generally smaller errors. In Scenario~\ref{scen:3}, \(\zeta=0.0001\) performs best, and the prediction errors increase for larger \(\zeta\).
These results highlight the sensitivity of prediction accuracy to the step size, motivating the wRV procedure developed in Section~\ref{subsec:adaptive-step}.
Figure~\ref{fig:beta_surface_comparison} displays the true slope surfaces \(\beta_0\) alongside their averaged online estimates at \(n=10000\), using \(\zeta=0.01\), \(0.0001\), and \(0.0001\) for Scenarios~\ref{scen:1}--\ref{scen:3}, respectively. These choices yield the (nearly) smallest prediction errors at \(n=10000\) in Table~\ref{tab:simulation_errors_zero}. 
The corresponding squared \(L^2\) estimation errors are reported in Table~S.1 in Section~S.3.1 of the Supplementary Material.
It turns out that the estimation error is also sensitive to the step size. Moreover, as with the prediction error, the estimation error also tends to become larger as the step size increases. 
\begin{table}[!ht]
\centering
\caption{
Prediction errors and Monte Carlo standard errors
(in parentheses) of the averaged online estimator with different
step sizes. All values are in units of $10^{-5}$.}
\label{tab:simulation_errors_zero}

\setlength{\tabcolsep}{2pt}
\renewcommand{\arraystretch}{1.05}

\resizebox{\textwidth}{!}{
\begin{tabular}{@{}cccccccc@{}}
\toprule
Scenario & $n$ & $\zeta=0.0001$ & $\zeta=0.001$ & $\zeta=0.01$ & $\zeta=0.05$ & $\zeta=0.1$ & $\zeta=0.2$ \\
\midrule

\multirow{4}{*}{1}
& 500 & 28.13(1.22) & 27.61(1.13) & 24.44(0.67) & \textbf{24.01}(0.37) & 28.27(0.30) & 40.11(0.31) \\
& 1000 & 12.33(0.29) & 12.21(0.28) & \textbf{11.72}(0.21) & 13.95(0.16) & 19.06(0.16) & 31.36(0.16) \\
& 5000 & 2.51(0.07) & 2.48(0.06) & \textbf{2.35}(0.04) & 2.91(0.03) & 4.57(0.04) & 12.85(0.06) \\
& 10000 & 1.39(0.03) & 1.37(0.03) & \textbf{1.32}(0.02) & 1.65(0.02) & 2.48(0.02) & 7.04(0.04) \\

\midrule

\multirow{4}{*}{2}
& 500 & \textbf{129.82}(1.18) & 130.16(1.14) & 134.97(0.95) & 165.68(0.81) & 207.20(0.72) & 280.45(0.56) \\
& 1000 & \textbf{70.60}(0.55) & 70.97(0.53) & 75.66(0.47) & 104.80(0.49) & 149.12(0.50) & 239.71(0.41) \\
& 5000 & \textbf{23.96}(0.06) & 23.97(0.06) & 24.28(0.06) & 28.18(0.07) & 40.29(0.12) & 106.59(0.20) \\
& 10000 & 21.95(0.03) & \textbf{21.93}(0.03) & 21.96(0.02) & 23.03(0.03) & 26.87(0.05) & 60.46(0.11) \\

\midrule

\multirow{4}{*}{3}
& 500 & \textbf{224.16}(1.98) & 224.81(1.89) & 233.55(1.38) & 290.43(1.06) & 370.41(1.01) & 557.13(1.41) \\
& 1000 & \textbf{121.63}(0.68) & 122.42(0.67) & 131.29(0.63) & 183.29(0.68) & 264.35(0.74) & 451.49(0.84) \\
& 5000 & \textbf{28.04}(0.20) & 28.22(0.20) & 30.29(0.16) & 43.02(0.14) & 69.23(0.18) & 190.55(0.32) \\
& 10000 & \textbf{12.75}(0.12) & 12.88(0.12) & 14.28(0.09) & 22.63(0.08) & 37.92(0.09) & 107.33(0.17) \\

\bottomrule
\end{tabular}
}
\end{table}

We then evaluate the practical performance of the wRV procedure. We set \(\xi=1\) in $ \operatorname{wRV}_{n,k}$ defined in \eqref{eq:wrv-criterion} and consider \(c_k\in\{0.85,0.925,1,1.075,1.15\}\), coupled with the base exponent \(\zeta=0.001\) in Scenario~\ref{scen:1} and \(\zeta=0.0001\) in Scenarios~\ref{scen:2} and \ref{scen:3}. As shown in the online wRV rows of Table~\ref{tab:simulation_errors_wrv}, the wRV-tuned estimators generally match or improve upon the best fixed-step prediction errors from Table~\ref{tab:simulation_errors_zero}. In Scenario~\ref{scen:1}, wRV produces modest reductions in prediction error across all sample sizes. In Scenario~\ref{scen:2}, the improvements are more visible at \(n=500\) and \(1000\). The improvement is most pronounced in Scenario~\ref{scen:3}, particularly for \(n=5000\) and \(10000\), indicating that rolling prediction losses provide an effective criterion for selecting the step size.

\begin{table}[!ht]
\centering
\normalsize
\setlength{\tabcolsep}{3pt}
\renewcommand{\arraystretch}{1.05}
\caption{
Prediction errors and Monte Carlo standard errors (in parentheses) of the wRV-tuned online estimator and the offline FLR estimator, reported in units of \(10^{-5}\).
}
\label{tab:simulation_errors_wrv}
\vspace{-3pt}
\begin{tabular*}{\textwidth}{@{\extracolsep{\fill}}cccccc@{}}
\toprule
Scenario & Method & \(n=500\) & \(n=1000\) & \(n=5000\) & \(n=10000\) \\
\midrule
\multirow{2}{*}{1}
& wRV     & 22.79\,(0.50) & 11.49\,(0.22) & 2.31\,(0.04) & 1.30\,(0.02) \\
& Offline & 15.03\,(0.57) & 8.13\,(0.27)  & 1.91\,(0.06) & 1.10\,(0.03) \\
\midrule
\multirow{2}{*}{2}
& wRV     & 123.78\,(1.40) & 65.46\,(0.68) & 23.92\,(0.07) & 21.93\,(0.03) \\
& Offline & 48.76\,(1.44)  & 39.09\,(0.84) & 22.30\,(0.45) & 13.43\,(0.24) \\
\midrule
\multirow{2}{*}{3}
& wRV     & 211.53\,(3.34) & 111.20\,(1.04) & 25.51\,(0.34) & 11.14\,(0.21) \\
& Offline & 35.92\,(1.28)  & 21.27\,(0.74) & 9.67\,(0.27)  & 8.65\,(0.20) \\
\bottomrule
\end{tabular*}
\end{table}

To systematically benchmark the statistical accuracy of our proposed method, Tables~\ref{tab:simulation_errors_wrv} and \ref{tab:simulation_abs_l2_offline} directly compare the prediction and squared \(L^2\) estimation errors of the wRV-tuned online estimator against the offline penalized RKHS estimator of \citet{sun2018optimal}, respectively. The offline procedure computes a global regularized estimate using the entire stored dataset and yields smaller mean errors across all scenarios considered. However, the performance gaps between the online and offline estimators diminish consistently as more observations are processed. 
A plausible reason is that the inaccurate initial value has a greater negative impact on the accuracy of the averaged estimator for relatively small $n$.
Furthermore, the wRV-tuned online estimator frequently exhibits smaller Monte Carlo standard errors compared to the offline benchmark, particularly in Scenarios~\ref{scen:1} and \ref{scen:2}, suggesting that our adaptive method can achieve satisfactory statistical accuracy, 
%with lower variability in prediction and estimation errors in some settings, 
while requiring only one pass through the data without storage of the full sample.

\begin{table}[!ht]
\centering
\normalsize
\setlength{\tabcolsep}{3pt}
\renewcommand{\arraystretch}{1.05}
\caption{
Squared \(L^2\) estimation errors and Monte Carlo standard errors (in parentheses) of the wRV-tuned online estimator and the offline FLR estimator, reported in units of \(10^{-5}\).
}
\label{tab:simulation_abs_l2_offline}
\vspace{-3pt}
\begin{tabular*}{\textwidth}{@{\extracolsep{\fill}}cccccc@{}}
\toprule
Scenario & Method & \(n=500\) & \(n=1000\) & \(n=5000\) & \(n=10000\) \\
\midrule
\multirow{2}{*}{1}
& wRV     & 72.85\,(1.37)  & 43.78\,(0.73)  & 13.13\,(0.25) & 7.53\,(0.16) \\
& Offline & 45.39\,(1.88)  & 27.02\,(0.94)  & 7.20\,(0.26)  & 4.33\,(0.14) \\

\midrule
\multirow{2}{*}{2}
& wRV     & 558.07\,(4.83) & 340.47\,(2.53) & 186.82\,(0.26) & 180.07\,(0.11) \\
& Offline & 257.35\,(6.88) & 236.20\,(6.11) & 172.30\,(3.80) & 114.88\,(2.60) \\
\midrule
\multirow{2}{*}{3}
& wRV     & 782.10\,(6.70) & 549.84\,(4.75) & 190.39\,(2.99) & 81.74\,(1.90) \\
& Offline & 120.29\,(6.54) & 76.58\,(4.53)  & 51.77\,(2.42) & 53.92\,(1.92) \\
\bottomrule
\end{tabular*}
\end{table}

\subsection{Confidence intervals for \texorpdfstring{\(\beta_0\)}{beta\_0}}\label{subsec:bootstrap_inference}

In this subsection, we evaluate the empirical performance of the proposed pointwise confidence intervals under Scenarios~\ref{scen:1}--\ref{scen:3}. 
The functional SGD and its bootstrap counterparts use the same step size \(\gamma_n=n^{-\zeta}\) in each run of sample size \(n\). Motivated by the inferential requirements of controlling stochastic remainders and initialization bias, we adopt \(\zeta = 0.35\) as a fixed numerical
choice. This choice yields satisfactory coverage in the main simulation study, while supplementary sensitivity analysis examines its performance relative to alternative exponents, as reported in Table~S.2 in Section~S.3.2.
Across \(M=200\) Monte Carlo replications, we generate \(B=500\) bootstrap paths using equal-probability multiplier weights \(\omega_i^b \in \{0, 2\}\) to construct 95\% pointwise B-Q confidence intervals at each grid point \(z=(t,s)\in\mathcal{T}\times\mathcal{S}\).

The spatial averages of the empirical coverage probabilities and the corresponding average interval lengths are summarized in Table~\ref{tab:simulation_pointwise_inference}. In general, the empirical coverage probabilities are close to the nominal level, with some variation across scenarios. In Scenario~\ref{scen:1}, the modest overcoverage observed at \(n=500\) diminishes as the sample size increases, and the coverage probabilities at \(n=3000\) and \(5000\) are close to \(0.95\). Scenario~\ref{scen:2} exhibits mild undercoverage at smaller sample sizes, and coverage approaching the nominal level at \(n=5000\), whereas Scenario~\ref{scen:3} remains moderately conservative across different sample sizes. Concurrently, the average interval lengths decrease substantially and monotonically with the sample size in every scenario. Thus, the intervals become increasingly precise as the sample size increases, while the empirical coverage remains close to the nominal level.

\begin{table}[!ht]
\centering
\caption{
Empirical coverage probabilities of the 95\% pointwise confidence intervals and average lengths(in parentheses) under different scenarios. 
}
\label{tab:simulation_pointwise_inference}
\renewcommand{\arraystretch}{1.15}
\setlength{\tabcolsep}{3pt}
\begin{tabular*}{\textwidth}{
    @{\extracolsep{\fill}}
    c
    cccc
    @{}
}
\toprule
& \multicolumn{4}{c}{Sample Size (\(n\))} \\
\cmidrule(lr){2-5}
Scenario & \(n=500\) & \(n=1000\) & \(n=3000\) & \(n=5000\) \\
\midrule
1 & 0.969\,(0.027) & 0.961\,(0.017) & 0.953\,(0.009) & 0.951\,(0.007) \\
2 & 0.942\,(0.030) & 0.936\,(0.020) & 0.944\,(0.011) & 0.955\,(0.008) \\
3 & 0.977\,(0.057) & 0.977\,(0.037) & 0.968\,(0.018) & 0.970\,(0.013) \\
\bottomrule
\end{tabular*}
\end{table}

\begin{figure}[!ht]
    \centering
    \includegraphics[width=1\textwidth]{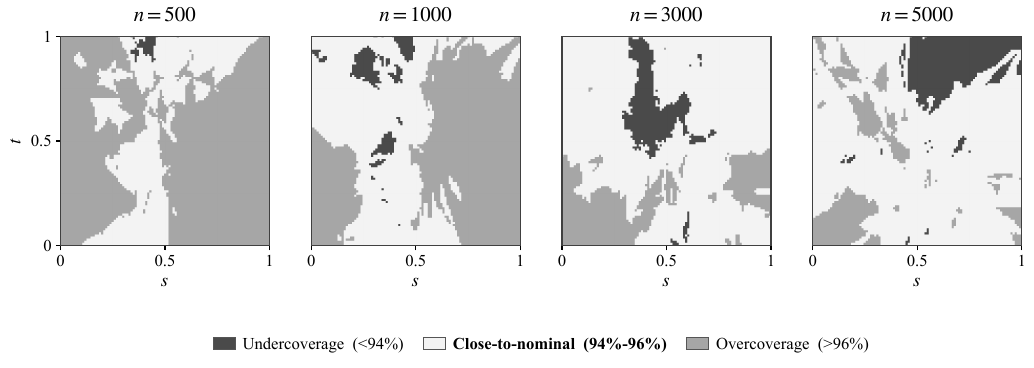}
    \caption{Spatial classification of the empirical pointwise coverage probabilities of the B-Q confidence intervals in Scenario~\ref{scen:1}.
    % for \(n=500\), \(1000\), \(3000\), and \(5000\).
    }
    \label{fig:bq_coverage}
\end{figure}

\begin{figure}[!ht]
    \centering
    \includegraphics[width=1\textwidth]{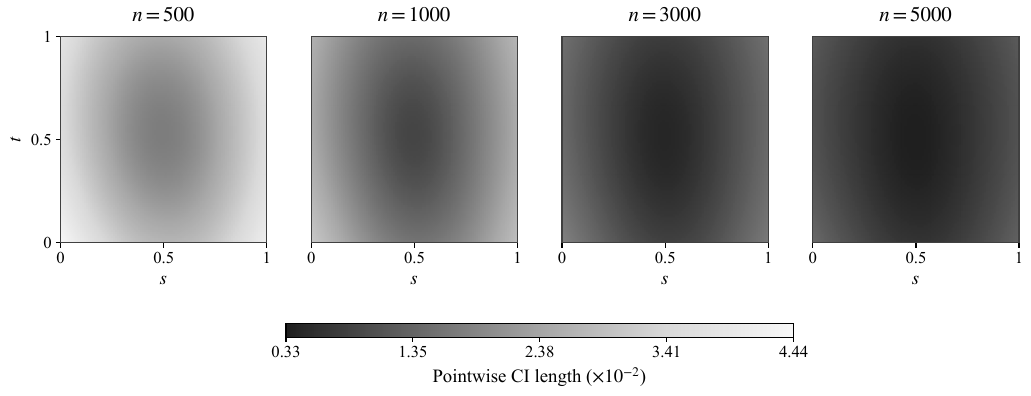}
    \caption{Average pointwise lengths of the B-Q confidence intervals in Scenario~\ref{scen:1}.
    }
    \label{fig:bq_length}
\end{figure}

Figures~\ref{fig:bq_coverage} and \ref{fig:bq_length} provide a more detailed view of the spatial resolution of the empirical pointwise coverage probabilities and average interval lengths for Scenario~\ref{scen:1}. In Figure~\ref{fig:bq_coverage}, the domain is classified into regions of undercoverage (below \(94\%\)), close-to-nominal coverage (\(94\%-96\%\)), and overcoverage (above \(96\%\)). While a substantial portion of the domain exhibits overcoverage at \(n=500\), this region shrinks considerably as the sample size increases, and a close-to-nominal coverage eventually dominates. In parallel, it can be seen in Figure~\ref{fig:bq_length} that the intervals are noticeably narrower in the interior of the domain and wider toward the boundaries. As \(n\) increases, these lengths shrink systematically across the entire \(\mathcal{T}\times\mathcal{S}\) region. Ultimately, these localized spatial visualizations provide direct graphical confirmation of the aggregated numerical summaries reported in Table~\ref{tab:simulation_pointwise_inference}. 
Corresponding spatial plots of empirical pointwise coverage probabilities and interval lengths for Scenarios~\ref{scen:2} and \ref{scen:3} are detailed in Figures~S.1--S.4 in Section~S.3.2 of the Supplementary Material.

\section{Real data examples}\label{sec:realdata}

We illustrate the proposed method through two real data applications. The first dataset contains a naturally ordered stream of daily ERA5 wind--wave trajectories, whereas the second examines the paired Chapman--Shaoxing electrocardiogram (ECG) beats randomly permuted into a sequential stream. In both applications, we aim to track how the estimated coefficient surface and its pointwise uncertainty evolve across nested stream lengths. For each stream length, the recursions are initialized at zero and fitted to the 
data prefix, with point wise B-Q confidence intervals constructed from \(B=500\) multiplier paths.

\subsection{Online wind-to-wave analysis using ERA5 reanalysis}\label{subsec:era5_wind_wave}

We apply the proposed online procedure to hourly ERA5 global reanalysis data \citep{hersbach2020era5}, which can be obtained from the \href{https://cds.climate.copernicus.eu/datasets/reanalysis-era5-single-levels-timeseries?tab=download}{Copernicus Climate Data Store} at an offshore North Sea location with coordinates \(56^\circ\mathrm{N}\) and \(3^\circ\mathrm{E}\). To investigate the one-day-lagged functional association between wind speed and ocean wave height,  
we define the functional predictor \(X_i(s)\) on day \(d_i\) as the 24-hour anomaly trajectory of the 10-m wind speed, derived from its orthogonal components as \(W = (u_{10}^{2} + v_{10}^{2})^{1/2}\), and specify the response \(Y_i(t)\) as the 24-hour significant wave height anomaly trajectory on the subsequent day \(d_i+1\). In addition, the discrete daily observations recorded at UTC hours \(h=0,\ldots,23\) are rescaled to ensure identical functional domains \(\mathcal{S}=\mathcal{T}=[0,1]\). We interpolate these hourly measurements using natural cubic splines and evaluate the resulting continuous curves at $61$ equally spaced grid points. To isolate the anomalies from the raw data spanning January 2010 through December 2025, we first verify timestamp completeness and remove non-finite or duplicate records, then eliminate recurring seasonal and diurnal variations by subtracting hour-specific, three-harmonic Fourier climatologies fitted exclusively to the 2010--2011 observations. Discarding this initial two-year estimation period yields a chronologically ordered data stream of $5113$ paired functional observations from January 2012 through December 2025.

\begin{figure}[!ht]
    \centering
    \includegraphics[width=1\textwidth]{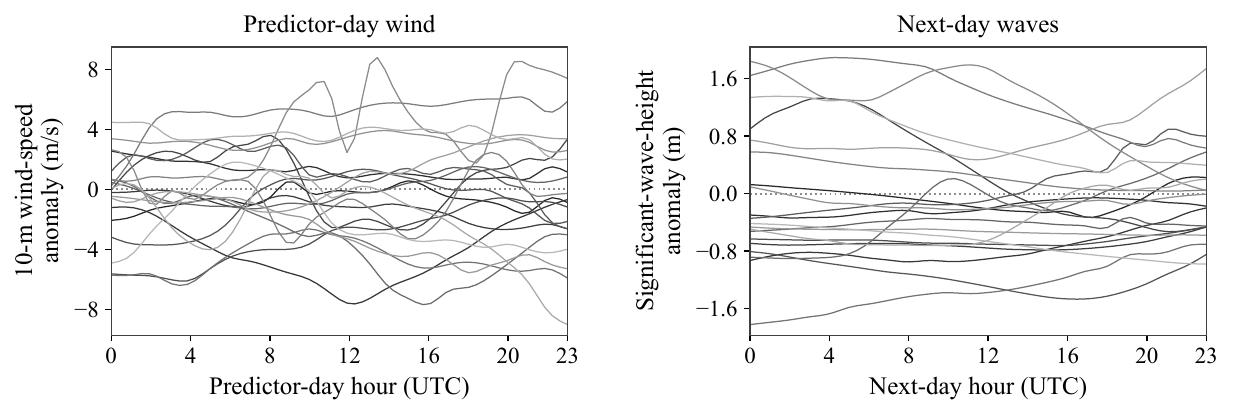}
    \caption{
    Twenty randomly selected paired ERA5 trajectories: seasonally adjusted predictor-day wind speed (left) and next-day significant wave height (right).
    }
    \label{fig:era5_random_curves}
\end{figure}

\begin{figure}[!ht]
    \centering
    \includegraphics[width=1\textwidth]{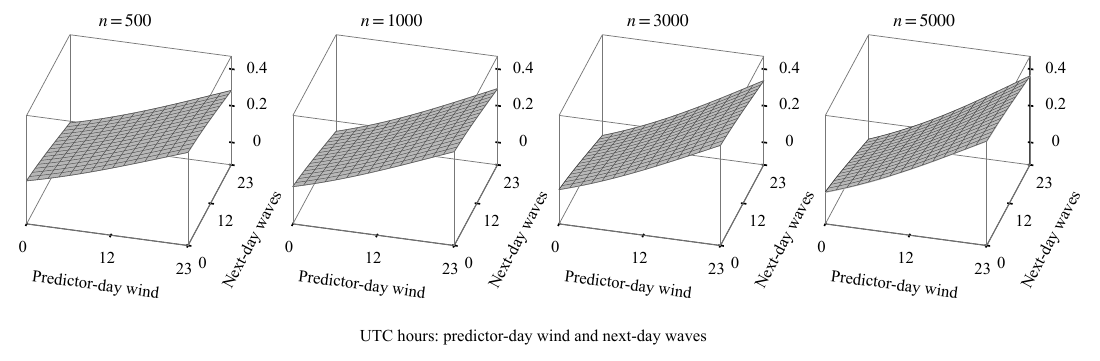}
    \caption{Estimated wind-to-wave coefficient surfaces for different stream lengths.
    % for \(n=500\), \(1000\), \(3000\), and \(5000\).
    }
    \label{fig:era5_coefficient_surface}
\end{figure}

Figure~\ref{fig:era5_random_curves} depicts appreciable day-to-day variation after seasonal adjustment. The wind-speed trajectories exhibit relatively pronounced intraday fluctuations, whereas the corresponding wave height trajectories tend to vary more smoothly over the day. We fit the proposed procedure to the first \(n=500\), \(1000\), \(3000\), and \(5000\) chronologically ordered pairs. Figure~\ref{fig:era5_coefficient_surface} presents the resulting coefficient estimates. The broad spatial pattern is similar across the four stream lengths. The estimated coefficient changes primarily along the
wind-speed time coordinate \(s\), i.e., negative or close to zero during the earlier portion of the predictor day and positive at later hours. By comparison, the surface varies less along the next-day time coordinate \(t\). Thus, under the fitted functional linear model, wind anomalies occurring later in the predictor day receive the largest positive weights in their association with the subsequent wave height anomaly trajectory. Although the estimates from the shorter prefixes exhibit greater variation in magnitude, the dominant direction of the association becomes increasingly stable as more observations are processed.

\begin{figure}[!ht]
    \centering
    \includegraphics[width=1\textwidth]{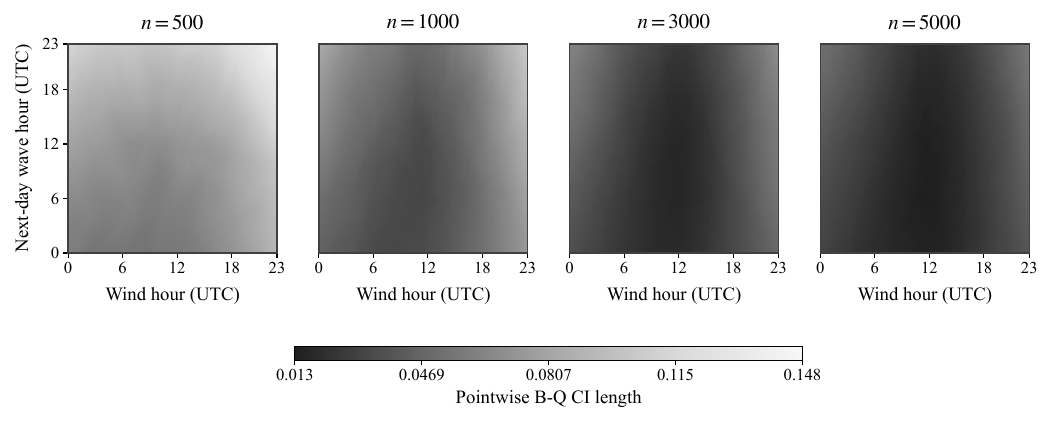}
    \caption{Lengths of the nominal \(95\%\) pointwise B-Q confidence intervals for the wind-to-wave coefficient surface. 
    % at \(n=500\), \(1000\), \(3000\), and \(5000\).
    }
    \label{fig:era5_CI_length}
\end{figure}

Figure~\ref{fig:era5_CI_length} displays the lengths of the corresponding pointwise B-Q confidence intervals. Under the common grayscale, the panels become progressively darker as \(n\) increases, indicating an overall reduction in pointwise uncertainty. The contraction is particularly pronounced between \(n=500\) and \(3000\), with a further reduction at \(n=5000\). The remaining spatial variation occurs mainly along the predictor-hour direction. The intervals are generally shorter over the interior of the wind-hour domain and longer near its boundaries, particularly near the late-hour boundary, whereas their variation across the wave-height response hour remains comparatively minimal.

\subsection{Online cross-lead analysis of 12-lead ECG recordings}\label{subsec:ecg}

The second application examines the association between simultaneously recorded ECG leads using a subset of 5000 recordings from a 12-lead database on \href{https://physionet.org/content/ecg-arrhythmia/1.0.0/}{PhysioNet} \citep{Zheng2020,PhysioNet-ecg-arrhythmia-1.0.0,pollard_physionet_2026}. We designate Lead~II as the functional predictor and Lead~V5 as the functional response. The \(500\)~Hz signals are filtered using a zero-phase \(50\)~Hz notch filter and a fourth-order Butterworth band-pass filter at \(0.5\)--\(40\)~Hz. Beats are extracted from \(0.30\) seconds before to \(0.50\) seconds after eligible Lead~II R peaks detected by WFDB XQRS. Beat screening uses peak-to-peak amplitude bounds and correlations with a within-record median template \citep{orphanidou2015signal}. Records with insufficient usable beats or median heart rates outside \(35\)--\(220\) beats per minute are excluded. Each retained record is represented by the pointwise median of its aligned beats \citep{Strodthoff2023}, without within-record amplitude standardization to preserve between-record amplitude variation. To facilitate functional data analysis, the representative trajectories are downsampled to a 101-point uniform grid. Let \(\theta\in[-0.30,0.50]\) denote the time in seconds relative to the Lead~II R-peak reference. We linearly map this physical interval to \([0,1]\) via \(u = (\theta + 0.30)/0.80\), positioning the reference at \(u=0.375\). This mapping yields the final paired functional observations \(X_i(s) = \widetilde V_{i,\mathrm{II}}(0.80s - 0.30)\) and \(Y_i(t) = \widetilde V_{i,\mathrm{V5}}(0.80t - 0.30)\) for \((s,t)\in[0,1]^2\), where \(\widetilde V\) denotes the representative voltage curve. Following this preprocessing pipeline, 4865 pairs of functional trajectories remain for the online analysis.

\begin{figure}[htbp]
    \centering
    \includegraphics[width=1\textwidth]{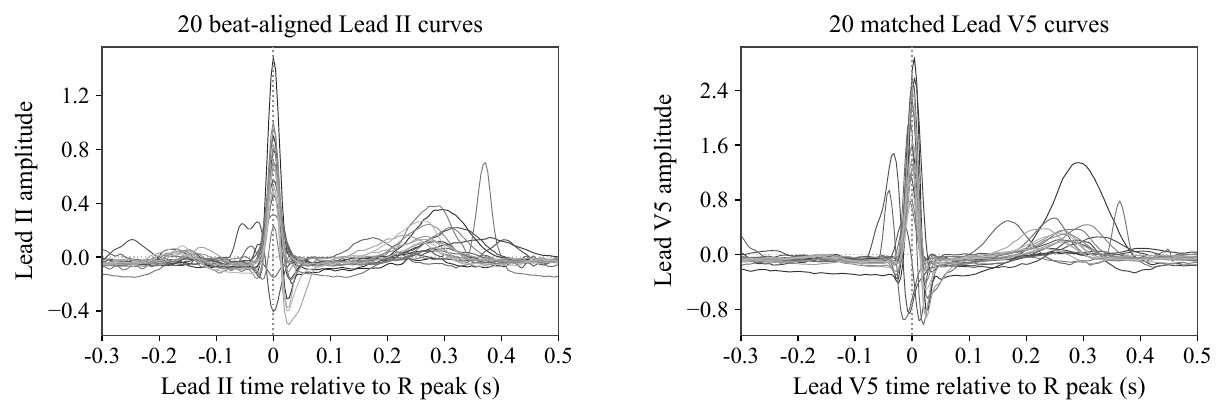}
    \caption{
    Twenty randomly selected paired representative ECG beats, aligned using Lead II R peaks: Lead II (left) and Lead V5 (right).
    }
    \label{fig:ecg_random_curves}
\end{figure}

Figure~\ref{fig:ecg_random_curves} illustrates the common Lead~II R-peak registration while showing substantial between-record variation in amplitude and morphology. Using the same alignment for both leads preserves their relative timing within each record. Since the records do not form a naturally ordered longitudinal sequence, we use one fixed random permutation (seed \(20260301\)) to define the stream and retain this ordering across all sample sizes.

\begin{figure}[!ht]
    \centering
    \includegraphics[width=1\textwidth]{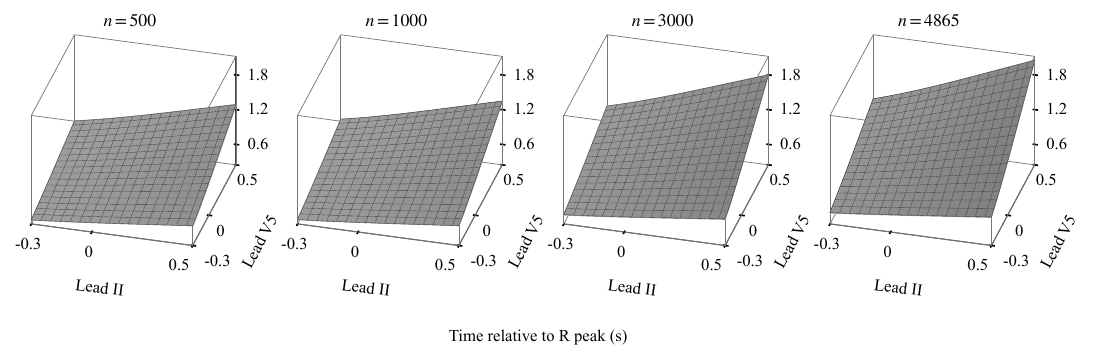}
    \caption{Estimated Lead II-to-Lead V5 coefficient surfaces for different stream lengths.
    }
    \label{fig:ecg_coefficient_surface}
\end{figure}

Figure~\ref{fig:ecg_coefficient_surface} shows that the estimated Lead II-to-Lead V5 coefficient is positive over the plotted domain at all four stream lengths. Its principal variation occurs along the response phase \(t\). Specifically, the estimated coefficient increases from the early to the late portion of the Lead V5 window, whereas its variation along the predictor phase \(s\) is more modest. The broad spatial pattern is already visible at \(n=500\), although the magnitude of the surface changes as more records are processed, particularly over the later response phases. The estimates obtained at \(n=3000\) and \(n=4865\) have similar overall shapes, suggesting that the dominant association pattern should be comparatively stable for a larger stream prefix. 

\begin{figure}[!ht]
    \centering
    \includegraphics[width=1\textwidth]{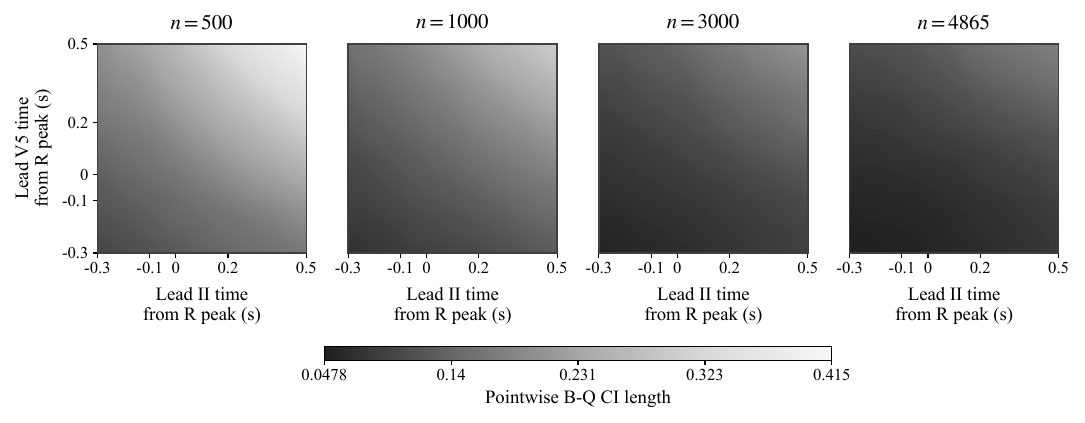}
    \caption{Lengths of the nominal \(95\%\) pointwise B-Q confidence intervals for the Lead II-to-Lead V5 coefficient surface.
    }
    \label{fig:ecg_CI_length}
\end{figure}

Likewise, The confidence interval lengths in Figure~\ref{fig:ecg_CI_length} decrease throughout the domain as the processed records increase. Unlike the ERA5 application, the spatial variation in the ECG interval lengths occurs mainly along the response-phase direction. The intervals are wider toward the later portion of the Lead V5 window and vary little comparatively with the Lead II phase. A larger stream prefix therefore improves pointwise precision throughout the domain while preserving a response-phase-dependent pattern of uncertainty.

\section{Conclusion}\label{sec:conclusion}

In this paper, we establish a one-pass estimation and pointwise inference framework for the streaming function-on-function linear regression. We estimate the bivariate coefficient surface in a tensor-product RKHS via  the proposed functional SGD with Polyak--Ruppert averaging, accompanied by an online multiplier bootstrap that avoids estimating the full covariance operator. Each observation can be discarded immediately after the update so that the 
storage does not increase with the length of the data stream. Theoretically, we establish convergence rates for the proposed functional SGD estimator, together with the conditional consistency of the bootstrap approximation for the pointwise distribution of the bias-centered estimation error. When the leading bias is asymptotically negligible, the proposed procedure provides pointwise inference for the coefficient surface itself. Numerical studies and applications to the ERA5 wind--wave and Chapman--Shaoxing ECG data demonstrate the finite-sample performance and practical utility of the proposed method. To the best of our knowledge, this is the first theoretically justified online bootstrap procedure for pointwise inference in streaming function-on-function regression.

Several directions warrant further investigation. First, while we currently focus on the function-on-function linear regression model, extending the proposed online estimation and inference framework to generalized or nonlinear models would be of interest. Second, the current inferential theory is pointwise and developed under independent sampling. Further investigation of simultaneous confidence bands and temporally dependent functional data streams would broaden the applicability of the proposed method. Finally, it would be useful to extend the theoretical analysis to genuinely adaptive step size schemes \citep{fang2023online}, beyond selecting among prespecified deterministic step size sequences.

\bibliographystyle{natbib.bst}
\bibliography{coding_note/bibliography}

\begin{thebibliography}{}

\bibitem[Bach and Moulines(2013)Bach and Moulines]{bach2013non}
Bach, F. and Moulines, E. (2013).
\newblock {Non-strongly-convex smooth stochastic approximation with convergence rate O (1/n)}.
\newblock {\em Advances in Neural Information Processing Systems\/}, {\bf 26}.

\bibitem[Billingsley(2017)Billingsley]{billingsley2017probability}
Billingsley, P. (2017).
\newblock {\em {Probability and Measure}\/}.
\newblock Wiley, New York.

\bibitem[Bottou {\em et~al.}(2018)Bottou, Curtis, and Nocedal]{bottou2018optimization}
Bottou, L., Curtis, F.~E., and Nocedal, J. (2018).
\newblock {Optimization Methods for Large-Scale Machine Learning}.
\newblock {\em SIAM Review\/}, {\bf 60}(2), 223--311.

\bibitem[Cai and Hall(2006)Cai and Hall]{CaiHall2006}
Cai, T.~T. and Hall, P. (2006).
\newblock {Prediction in Functional Linear Regression}.
\newblock {\em The Annals of Statistics\/}, {\bf 34}(5), 2159--2179.

\bibitem[Cai and Yuan(2012)Cai and Yuan]{Cai01092012}
Cai, T.~T. and Yuan, M. (2012).
\newblock {Minimax and Adaptive Prediction for Functional Linear Regression}.
\newblock {\em Journal of the American Statistical Association\/}, {\bf 107}(499), 1201--1216.

\bibitem[Cardot {\em et~al.}(2003)Cardot, Ferraty, and Sarda]{CardotFerratySarda2003}
Cardot, H., Ferraty, F., and Sarda, P. (2003).
\newblock {Spline Estimators for the Functional Linear Model}.
\newblock {\em Statistica Sinica\/}, {\bf 13}(3), 571--591.

\bibitem[Chen {\em et~al.}(2022)Chen, Tang, Fan, and Guo]{ChenTangFanGuo2022}
Chen, X., Tang, B., Fan, J., and Guo, X. (2022).
\newblock {Online Gradient Descent Algorithms for Functional Data Learning}.
\newblock {\em Journal of Complexity\/}, {\bf 70}, 101635.

\bibitem[Cheng {\em et~al.}(2025)Cheng, Hu, Lin, and Wang]{cheng2025online}
Cheng, G., Hu, W., Lin, R., and Wang, C. (2025).
\newblock {Online Robust Estimation and Bootstrap Inference for Function-on-Scalar Regression}.
\newblock {\em Statistics and Computing\/}, {\bf 35}(1), 13.

\bibitem[Crambes and Mas(2013)Crambes and Mas]{CrambesMas2013}
Crambes, C. and Mas, A. (2013).
\newblock {Asymptotics of Prediction in Functional Linear Regression with Functional Outputs}.
\newblock {\em Bernoulli\/}, {\bf 19}(5B), 2627--2651.

\bibitem[Cucker and Smale(2002)Cucker and Smale]{cucker2002mathematical}
Cucker, F. and Smale, S. (2002).
\newblock {On the Mathematical Foundations of Learning}.
\newblock {\em Bulletin of the American Mathematical Society\/}, {\bf 39}(1), 1--49.

\bibitem[Cuevas {\em et~al.}(2002)Cuevas, Febrero, and Fraiman]{cuevas2002linear}
Cuevas, A., Febrero, M., and Fraiman, R. (2002).
\newblock {Linear Functional Regression: The Case of Fixed Design and Functional Response}.
\newblock {\em Canadian Journal of Statistics\/}, {\bf 30}(2), 285--300.

\bibitem[Dette and Tang(2024)Dette and Tang]{dette2024statistical}
Dette, H. and Tang, J. (2024).
\newblock {Statistical Inference for Function-on-Function Linear Regression}.
\newblock {\em Bernoulli\/}, {\bf 30}(1), 304--331.

\bibitem[Dieuleveut and Bach(2016)Dieuleveut and Bach]{dieuleveut2016nonparametric}
Dieuleveut, A. and Bach, F. (2016).
\newblock {Nonparametric Stochastic Approximation with Large Step-Sizes}.
\newblock {\em The Annals of Statistics\/}, {\bf 44}(4), 1363--1399.

\bibitem[Fang {\em et~al.}(2018)Fang, Xu, and Yang]{fang2018online}
Fang, Y., Xu, J., and Yang, L. (2018).
\newblock {Online Bootstrap Confidence Intervals for the Stochastic Gradient Descent Estimator}.
\newblock {\em Journal of Machine Learning Research\/}, {\bf 19}(78), 1--21.

\bibitem[Gu(2013)Gu]{gu2013}
Gu, C. (2013).
\newblock {\em Smoothing {Spline ANOVA Models}\/}.
\newblock Springer, New York, 2 edition.

\bibitem[Hall and Horowitz(2007)Hall and Horowitz]{HallHorowitz2007}
Hall, P. and Horowitz, J.~L. (2007).
\newblock {Methodology and Convergence Rates for Functional Linear Regression}.
\newblock {\em The Annals of Statistics\/}, {\bf 35}(1), 70--91.

\bibitem[Hersbach {\em et~al.}(2020)Hersbach, Bell, Berrisford, Hirahara, Horányi, Muñoz-Sabater, {\em et~al.}]{hersbach2020era5}
Hersbach, H., Bell, B., Berrisford, P., Hirahara, S., Horányi, A., Muñoz-Sabater, J., {\em et~al.} (2020).
\newblock {The ERA5 Global Reanalysis}.
\newblock {\em Quarterly Journal of the Royal Meteorological Society\/}, {\bf 146}(730), 1999--2049.

\bibitem[Hsing and Eubank(2015)Hsing and Eubank]{hsing2015theoretical}
Hsing, T. and Eubank, R.~L. (2015).
\newblock {\em {Theoretical Foundations of Functional Data Analysis, with An Introduction to Linear Operators}\/}.
\newblock Wiley, New York.

\bibitem[Imaizumi and Kato(2018)Imaizumi and Kato]{imaizumi2018pca}
Imaizumi, M. and Kato, K. (2018).
\newblock {PCA-Based Estimation for Functional Linear Regression with Functional Responses}.
\newblock {\em Journal of Multivariate Analysis\/}, {\bf 163}, 15--36.

\bibitem[Kokoszka and Reimherr(2017)Kokoszka and Reimherr]{kokoszka2017introduction}
Kokoszka, P. and Reimherr, M. (2017).
\newblock {\em {Introduction to Functional Data Analysis}\/}.
\newblock Chapman and Hall/CRC, New York.

\bibitem[Lian(2015)Lian]{LIAN2015395}
Lian, H. (2015).
\newblock {Minimax Prediction for Functional Linear Regression with Functional Responses in Reproducing Kernel Hilbert Spaces}.
\newblock {\em {Journal of Multivariate Analysis}\/}, {\bf 140}, 395--402.

\bibitem[Luo and Qi(2017)Luo and Qi]{LuoQi2017}
Luo, R. and Qi, X. (2017).
\newblock {Function-on-Function Linear Regression by Signal Compression}.
\newblock {\em Journal of the American Statistical Association\/}, {\bf 112}(518), 690--705.

\bibitem[M{\"u}cke {\em et~al.}(2019)M{\"u}cke, Neu, and Rosasco]{mucke2019beating}
M{\"u}cke, N., Neu, G., and Rosasco, L. (2019).
\newblock {Beating SGD saturation with tail-averaging and minibatching}.
\newblock {\em Advances in Neural Information Processing Systems\/}, {\bf 32}.

\bibitem[Orphanidou {\em et~al.}(2015)Orphanidou, Bonnici, Charlton, Clifton, Vallance, and Tarassenko]{orphanidou2015signal}
Orphanidou, C., Bonnici, T., Charlton, P., Clifton, D., Vallance, D., and Tarassenko, L. (2015).
\newblock {Signal-Quality Indices for the Electrocardiogram and Photoplethysmogram: Derivation and Applications to Wireless Monitoring}.
\newblock {\em IEEE Journal of Biomedical and Health Informatics\/}, {\bf 19}(3), 832--838.

\bibitem[Pollard {\em et~al.}(2026)Pollard, Moody, Lehman, Gow, Fernandes, Xie, Johnson, Mark, and Heldt]{pollard_physionet_2026}
Pollard, T., Moody, B.~E., Lehman, L.-w.~H., Gow, B.~J., Fernandes, C., Xie, C., Johnson, A., Mark, R.~G., and Heldt, T. (2026).
\newblock {PhysioNet as a Global Platform for Biomedical Research}.
\newblock {\em Nature Health\/}, {\bf 1}(8), 792--795.

\bibitem[Polyak and Juditsky(1992)Polyak and Juditsky]{polyak1992acceleration}
Polyak, B.~T. and Juditsky, A.~B. (1992).
\newblock {Acceleration of Stochastic Approximation by Averaging}.
\newblock {\em SIAM Journal on Control and Optimization\/}, {\bf 30}(4), 838--855.

\bibitem[Ramsay and Dalzell(1991)Ramsay and Dalzell]{RamsayDalzell1991}
Ramsay, J.~O. and Dalzell, C.~J. (1991).
\newblock {Some Tools for Functional Data Analysis}.
\newblock {\em Journal of the Royal Statistical Society: Series B (Methodological)\/}, {\bf 53}(3), 539--561.

\bibitem[Ramsay and Silverman(2005)Ramsay and Silverman]{Ramsay1997}
Ramsay, J.~O. and Silverman, B.~W. (2005).
\newblock {\em {Functional Data Analysis}\/}.
\newblock Springer, New York, 2 edition.

\bibitem[Robbins and Monro(1951)Robbins and Monro]{robbins1951stochastic}
Robbins, H. and Monro, S. (1951).
\newblock {A Stochastic Approximation Method}.
\newblock {\em The Annals of Mathematical Statistics\/}, {\bf 22}(3), 400--407.

\bibitem[Ruppert(1988)Ruppert]{ruppert1988efficient}
Ruppert, D. (1988).
\newblock {Efficient Estimations from a Slowly Convergent Robbins-Monro Process}.
\newblock Technical report, Cornell University Operations Research and Industrial Engineering.

\bibitem[Strodthoff {\em et~al.}(2023)Strodthoff, Mehari, Nagel, Aston, Sundar, Graff, Kanters, Haverkamp, D{\"o}ssel, Loewe, B{\"a}r, and Schaeffter]{Strodthoff2023}
Strodthoff, N., Mehari, T., Nagel, C., Aston, P.~J., Sundar, A., Graff, C., Kanters, J.~K., Haverkamp, W., D{\"o}ssel, O., Loewe, A., B{\"a}r, M., and Schaeffter, T. (2023).
\newblock {{PTB-XL+}, a Comprehensive Electrocardiographic Feature dataset}.
\newblock {\em Scientific Data\/}, {\bf 10}(1), 279.

\bibitem[Sun {\em et~al.}(2018)Sun, Du, Wang, and Ma]{sun2018optimal}
Sun, X., Du, P., Wang, X., and Ma, P. (2018).
\newblock {Optimal Penalized Function-on-Function Regression under a Reproducing Kernel Hilbert Space Framework}.
\newblock {\em Journal of the American Statistical Association\/}, {\bf 113}(524), 1601--1611.

\bibitem[Wang {\em et~al.}(2016)Wang, Chiou, and M{\"u}ller]{wang2016functional}
Wang, J.-L., Chiou, J.-M., and M{\"u}ller, H.-G. (2016).
\newblock {Functional Data Analysis}.
\newblock {\em Annual Review of Statistics and Its Application\/}, {\bf 3}, 257--295.

\bibitem[Xie {\em et~al.}(2026)Xie, Shi, Sang, Shang, Jiang, and Kong]{xie2026scalable}
Xie, J., Shi, E., Sang, P., Shang, Z., Jiang, B., and Kong, L. (2026).
\newblock {Scalable Inference in Functional Linear Regression with Streaming Data}.
\newblock {\em The Annals of Statistics\/}, {\bf 54}(1), 120--145.

\bibitem[Yan and Yao(2026)Yan and Yao]{yan2026functional}
Yan, S. and Yao, F. (2026).
\newblock {Functional Linear Regression from Sparse to Dense Designs: A Pooling-ridge Method and Minimax Optimality}.
\newblock {\em The Annals of Statistics\/}.
\newblock To appear.

\bibitem[Yang and Yao(2023)Yang and Yao]{fang2023online}
Yang, Y. and Yao, F. (2023).
\newblock {Online Estimation for Functional Data}.
\newblock {\em Journal of the American Statistical Association\/}, {\bf 118}(543), 1630--1644.

\bibitem[Yao {\em et~al.}(2005)Yao, M{\"u}ller, and Wang]{yao2005functional}
Yao, F., M{\"u}ller, H.-G., and Wang, J.-L. (2005).
\newblock {Functional Linear Regression Analysis for Longitudinal Data}.
\newblock {\em The Annals of Statistics\/}, {\bf 33}(6), 2873--2903.

\bibitem[Yuan and Cai(2010)Yuan and Cai]{yuan2010reproducing}
Yuan, M. and Cai, T.~T. (2010).
\newblock {A Reproducing Kernel Hilbert Space Approach to Functional Linear Regression}.
\newblock {\em The Annals of Statistics\/}, {\bf 38}(6), 3412--3444.

\bibitem[Zhang and Lei(2025)Zhang and Lei]{ZhangLei2025}
Zhang, T. and Lei, J. (2025).
\newblock {Online Estimation with Rolling Validation: Adaptive Nonparametric Estimation with Streaming Data}.
\newblock {\em The Annals of Statistics\/}, {\bf 53}(6), 2702--2727.

\bibitem[Zheng {\em et~al.}(2020)Zheng, Chu, Struppa, Zhang, Yacoub, El-Askary, {\em et~al.}]{Zheng2020}
Zheng, J., Chu, H., Struppa, D., Zhang, J., Yacoub, M., El-Askary, H., {\em et~al.} (2020).
\newblock {Optimal Multi-Stage Arrhythmia Classification Approach}.
\newblock {\em Scientific Reports\/}, {\bf 10}, 2898.

\bibitem[Zheng {\em et~al.}(2022)Zheng, Guo, and Chu]{PhysioNet-ecg-arrhythmia-1.0.0}
Zheng, J., Guo, H., and Chu, H. (2022).
\newblock {A Large Scale 12-Lead Electrocardiogram Database for Arrhythmia Study}.
\newblock {\em {PhysioNet}\/}.
\newblock Version 1.0.0.

\end{thebibliography}
\end{document}